\documentstyle[11pt]{article}

\newcommand{\be}{\begin{equation}}
\newcommand{\ee}{\end{equation}}
\newcommand{\nn}{\nonumber}
\newcommand{\beba}{\begin{equation}\begin{array}{lcl}}
\newcommand{\eaee}{\end{array}\end{equation}}
\newcommand{\bea}{\begin{eqnarray}}
\newcommand{\eea}{\end{eqnarray}}
\newcommand{\ba}{\begin{array}}
\newcommand{\ea}{\end{array}}

\newcommand{\ns}{\normalsize}
\newcommand{\refs}[1]{(\ref{#1})}

\def\a{\alpha}
\def\b{\beta}
\def\g{\gamma}

\def\d{\delta}
\def\e{\epsilon}

\def\k{\kappa}

\def\m{\mu}
\def\n{\nu}

\def\r{\rho}
\def\s{\sigma}

\def\x{\xi}

\def\F{\Phi}

\def\wtd{\widetilde}

\def\DH{\hat{D}}
\def\gh{\hat{g}}
\def\bh{\hat{b}}
\def\sg{\sqrt{-\g}}
\def\pa{\partial}
\def\gh{\hat{g}}
\def\bh{\hat{b}}
\def\mb{{\bar{m}}}
\def\nb{{\bar{n}}}
\def\rb{{\bar{r}}}
\def\sb{{\bar{s}}}
\def\ght{\hat{g}\kern-0.6em \widetilde{\raisebox{-0.12em}{\phantom{X}}}}
\def\bht{\hat{b}\kern-0.6em \widetilde{\raisebox{0.15em}{\phantom{X}}}}
\def\cF{{\cal F}}
\def\bG{{\bf G}}

\begin{document}


\begin{titlepage}
\title{\hfill{\ns UPR-747T, IASSNS-HEP-97/42\\}
       \hfill{\ns hep-th/9704178\\[.1cm]}
       \hfill{\ns April 1997}\\[.8cm]
       {\Large\bf U--Duality Symmetries from the Membrane Worldvolume}}
\author{Andr\'e
        Lukas$^1$\setcounter{footnote}{0}\thanks{Supported by Deutsche
        Forschungsgemeinschaft (DFG) and
        Nato Collaborative Research Grant CRG.~940784.}~~and
        Burt A.~Ovrut$^1\; ^2$\\[0.5cm]
        {\ns $^1$Department of Physics, University of Pennsylvania} \\
        {\ns Philadelphia, PA 19104--6396, USA}\\[0.3cm]
        {\ns $^2$School of Natural Sciences, Institute for Advanced Study}\\
        {\ns Olden Lane, Princeton, NJ 08540, USA}}
\date{}
\maketitle

\begin{abstract}
We study U--duality symmetries of toroidally compactified M theory from
the membrane worldvolume point of view. This is done taking the most
general set of bosonic background fields into account. Upon restriction to pure
moduli backgrounds, we are able to find the correct U--duality groups
and moduli coset parameterizations for dimensions $D>6$ as symmetries
of the membrane worldvolume theory. In particular,
we derive the $D=8$ U--duality group $SL(2)\times SL(3)$. For
general background fields, we concentrate on the case $D=8$. Though the
$SL(2)$ part of the symmetry appears to be obstructed by certain terms
in the equations of motion, we are able to read off the transformation
properties for the background fields. These transformations are verified
by comparison with 11--dimensional supergravity dimensionally reduced to
$D=8$.
\end{abstract}

\thispagestyle{empty}
\end{titlepage}


\section{Introduction}

Recently, string theory has undergone a dramatic evolution. It is now
believed that the perturbative expansions of the five consistent string
theories represent patches in the moduli space of a
yet larger theory, called M--theory~\cite{tow_rep}. The low energy limit
of this theory is given by 11--dimensional supergravity. Though the correct
quantization of M theory is presently unknown, it is believed to be a theory
of fundamental membranes. Correspondingly, the starting point for a
quantization of M theory~\cite{quant} is the 11--dimensional supermembrane
action~\cite{hlp}. Classically, imposing $\k$ symmetry on the supermembrane
worldvolume restricts the background fields, the metric,
a 3--form field and their superpartners, to fulfill the equations of
motion of 11--dimensional supergravity~\cite{kappa}.

Duality symmetries play an important r\^ole in this picture in that they
relate the various string theories and 11--dimensional supergravity,
compactified on different backgrounds, to each other. This provides evidence
for the existence of a more general underlying theory.  These relations have
basically been established by showing that the soliton
spectra~\cite{duff_sol} of certain theories are mapped into each other
by duality transformations. Clearly, a major goal is to prove the invariance
of the underlying theory, compactified quantum M theory, under these
symmetries. A first step toward such a proof is to analyze to what extent these
symmetries are realized in the corresponding classical theory, namely
in the 11--dimensional supermembrane. It is this question which we are
going to address in the present paper, for the case of M theory
toroidally compactified to $D$ space--time dimensions. The corresponding
U--duality symmetries have been known for a long time as classical continuous
symmetries of dimensionally reduced 11--dimensional supergravity~\cite{cj}.
Recently, Hull and Townsend~\cite{ht} found evidence that discrete versions
of these continuous U--duality groups are symmetries of the full theory,
by showing the invariance of certain BPS soliton spectra. Since these
U--duality symmetries include T as well as S duality, they provide a
``unified'' picture of dualities. It is the main purpose of this paper
to investigate to what extent this unified picture of dualities arises from
the worldvolume theory of the 11--dimensional supermembrane. Before
explaining this in more detail, however, let us briefly describe the
analogous problem in string theory.

\vspace{0.4cm}

T duality~\cite{T_dual} is a well known exact symmetry of perturbative
string theory. It can be seen to leave the perturbative spectrum invariant,
by exchanging momentum and winding modes, if simultaneously the
``radius of compactification'' is inverted. Classically, the discrete
T--duality symmetry enlarges to a continuous group of T--duality rotations
which arise as symmetries of the string $\s$ model. For a pure moduli
background, this has been shown in ref.~\cite{duff1} by analyzing the
symmetries of the combined system of equations of motion and Bianchi
identities. This applies a method developed earlier in the context of
4--dimensional gauge theories~\cite{gz} to the string $\s$ model.
The generalization to include the full bosonic background field content of the
theory was provided in ref.~\cite{ms}. Duality rotations on the
full background field content constitute a generalization of the
discrete Buscher duality transformation~\cite{buscher}.
On the other hand, S duality, as a consequence of string--string duality
in six dimensions~\cite{str_str}, is a symmetry of string theory in
$D=4$~\cite{S4,duff_rep}. Since S duality inverts the coupling constant of the
theory (the dilaton), it is a nonperturbative symmetry. As such, it does not
leave the perturbative spectrum of the theory invariant and its continuous
version is, therefore, not a classical symmetry of the string $\s$ model.
Instead, there is evidence that the r\^ole of T and S duality is exchanged
for the 5 brane (which is dual to the string in $D=10$) and that string
S duality can be discovered as a symmetry of the 5 brane worldvolume
theory~\cite{ss}.

The situation is quite different for M theory. Since, in this theory, the
dilaton is just a geometrical modulus (associated with the radius of the
compact eleventh dimension), one does not expect a conceptual difference
between S and T duality. Consequently, one may hope to find both symmetries,
and even the larger U--duality symmetry, from a single worldvolume theory;
namely, the $\DH = 11$ supermembrane. Then, formally, U duality would arise
from the supermembrane in the same way that T duality arises from the string
worldsheet~: either as a
classical continuous symmetry rotating equations of motion and Bianchi
identities into each other or as a discrete symmetry leaving the ``spectrum''
of the membrane invariant. For both pictures, there exists some evidence in
the literature. In a very interesting paper~\cite{dl}, Duff and Lu analyzed the
moduli part of the membrane equations of motion and showed explicitly the
appearance of the $D=7$ U--duality group $SL(5)$. The present paper is
highly motivated by this work. U duality as a symmetry of the membrane
spectrum has been discussed by Sen~\cite{sen}. He pointed out that, unlike
for the string, the dimension $d$ of the momentum vector (for a membrane
compactified to $D=11-d$ dimensions) generally does not equal the
dimension of the winding vector which is
$\frac{d(d-1)}{2}$. For $D=8$ only both dimensions coincide, so that there is
a close analogy to the string in this dimension. In this case, Sen argued
for the existence of an $SL(2)$ symmetry which leaves the theory invariant
by exchanging momentum and winding numbers of the membrane and simultaneously
transforming the dilaton and the 3 form modulus.

\vspace{0.4cm}

In this paper, we will analyze U duality rotations from the viewpoint
of the classical membrane worldvolume theory, following the method
of ref.~\cite{dl}. Our work differs from ref.~\cite{dl} in two important
aspects. First, we are considering the full 11--dimensional membrane
target space, which appears to be crucial for the interpretation of
the worldvolume symmetries as U--duality symmetries. Second, we include
the full spectrum of bosonic background fields in our analysis. This means that
we are looking for a continuous version of Buscher duality for the
membrane. The outline of the paper is as follows. In the next section,
we present a review of T duality rotations on the string worldsheet to
explain the methods which we will apply to the membrane later on.
Section 3 presents the general analysis for the membrane which follows
the string analog as closely as possible. In section 4, we concentrate
on pure moduli background field configurations and discuss all
cases with $D\geq 6$. In particular, we reproduce the $D=7$ result of Duff
and Lu and show how the $SL(2)$ symmetry of Sen arises within our setting.
The general case, including all background field, is discussed in section 5
for $D=8$. Although we find that the $SL(2)$ part of the
$D=8$ U--duality group is obstructed as a worldvolume symmetry, we are
able to read off the exact $SL(2)\times SL(3)$ transformation laws for all
fields except for the 3 form. These results are verified in section 6 by
comparison with $\DH =11$ supergravity dimensionally reduced to $D=8$. Finally,
in section 7 we summarize and comment on our results.

\section{Review of T Duality Rotations on the String Worldsheet}

In this section we will discuss how T duality rotations of a toroidally
compactified string arise as symmetries of the worldsheet $\s$--model.
The basic method which we will use is the analysis of rotations between
equations of motion and Bianchi identities as first discussed by Gaillard and
Zumino~\cite{gz} in the context of 4--dimensional gauge theories. This method
has been applied to the moduli part of the string worldsheet action by 
Duff~\cite{duff1} and we will follow the method of this paper.
The generalization to the full background field content
has been provided by Maharana and Schwarz~\cite{ms}.
Our main intention is to explain some of the methods, which we will later on
apply to the membrane worldvolume theory, within the more familiar setting
of string theory. Consequently, we will keep our presentation as close as
possible to the membrane case. In particular, unlike in ref.~\cite{ms}, we
will not work in conformal gauge but keep a general worldsheet metric.
Furthermore, since we wish to discuss the full content of background  
fields for the membrane, we will do so for the string as well.

We consider a string in $\DH =10$ dimensions with worldsheet coordinates
$\x^i$, $i,j,k,... =0,1$, worldsheet metric $\g_{ij}$ and target space
coordinates $X^M(\x^i )$. The full target space is indexed by uppercase
letters  $M,N,P,... = 0,...,\DH -1$. Its motion in the background
specified by the metric $\gh_{MN}=\gh_{MN}(X^P )$ and the antisymmetric
tensor field $\bh_{MN}=\bh_{MN}(X^P)$ is described by the $\s$--model
Lagrangian~\footnote{Our definition of the $\e$ symbol is such that
$\e^{01}=1$.}
\be
 {\cal L} = \frac{1}{2}\sg\g^{ij}\pa_iX^M\pa_jX^N\gh_{MN}
            +\frac{1}{2}\e^{ij}\pa_iX^M\pa_jX^N\bh_{MN}\; .
 \label{string}
\ee
The equation of motion for $\g^{ij}$ implies the vanishing of the 
energy momentum $T_{ij}$ tensor and reads
\be
 T_{ij}=\frac{1}{\sg}\frac{\pa{\cal L}}{\pa\g^{ij}}
       =\frac{1}{2}\left( \pa_iX^M\pa_jX^N\gh_{MN}-\g_{ij}\g^{kl}
        \pa_kX^M\pa_lX^N\gh_{MN}\right) = 0\; .\label{st_T}
\ee
Next we would like to dimensionally reduce the above Lagrangian to
$D=\DH -d$ dimensions. We split the target space coordinates
into an external and internal piece as $X^M =(X^\m ,X^m )$ and assume
that the background depends on the external coordinates only; that is,
$\gh_{MN} = \gh_{MN}(X^\m )$ and $\bh_{MN} = \bh_{MN}(X^\m )$. Indices
$\m ,\n ,\r  ,...= 0,...,D-1$ and $m,n,r,... = D,...,\DH -1$ are used to
denote the external and internal coordinates respectively. The background
fields are split in a corresponding way as
$\gh_{MN}=(\gh_{\m\n},\gh_{\m n},\gh_{mn})$
and $\bh_{MN}=(\bh_{\m\n},\bh_{\m n},\bh_{mn})$ where $\gh_{\m\n}$,
$\bh_{\m\n}$ are the external metric and antisymmetric tensor field,
$\gh_{\m n}$, $\bh_{\m n}$ represent $2d$ vector fields on the external space
and $\gh_{mn}$, $\bh_{mn}$ are the $d^2$ moduli of the internal space. The
Lagrangian~\refs{string} then decomposes into an external part, a mixed part
and an internal, pure moduli part as
\be
 {\cal L} = {\cal L}^{(0)}+{\cal L}^{(1)}+{\cal L}^{(2)}
\ee
with
\bea
 {\cal L}^{(0)} &=& \frac{1}{2}\sg\g^{ij}\pa_iX^\m\pa_jX^\n \gh_{\m\n}
            +\frac{1}{2}\e^{ij}\pa_iX^\m\pa_jX^\n \bh_{\m\n} \nn \\ 
 {\cal L}^{(1)} &=& \sg\g^{ij}\pa_iX^\m\pa_jX^n \gh_{\m n}
            +\e^{ij}\pa_iX^\m\pa_jX^n \bh_{\m n}
            \label{string_split}\\
 {\cal L}^{(2)} &=& \frac{1}{2}\sg\g^{ij}\pa_iX^m\pa_jX^n \gh_{mn}
            +\frac{1}{2}\e^{ij}\pa_iX^m\pa_jX^n \bh_{mn} \nn \; .
\eea
Since ${\cal L}$ depends on the coordinates $X^m$ through their
derivatives only, we may introduce ``field strengths'' $F_i^m$ and
rewrite ${\cal L}$ in an equivalent first order form as
\be
 {\cal L}_x = \left. {\cal L}^{(0)}\right|_{\pa X=F}-\left. {\cal L}^{(2)}
              \right|_{\pa X=F}+\left. \frac{\pa{\cal L}}
              {\pa (\pa_i X^m)}\right|_{\pa X=F}\pa_i X^m\; ,
 \label{st_Lx}
\ee
where $|_{\pa X =F}$ after an expression indicates that $\pa_iX^m$ has been
replaced by $F_i^m$. The equivalence of this first
order Lagrangian to the original one can be easily proven by looking
at the equation of motion for $F_i^m$,
\be
 \frac{\pa{\cal L}_x}{\pa F_i^m} = (\sg\g^{ij}\gh_{mn}+\e^{ij}\bh_{mn})
                                 (\pa_iX^m - F_i^m) = 0
\ee
which leads to $F_i^m=\pa_iX^m$. Substituting this solution into
${\cal L}_x$ leads back to ${\cal L}$. That is
\be
 {\cal L} = \left. {\cal L}_x\right|_{F=\pa X}\; . \label{st_LLx}
\ee
Consequently, the equations of motion for ${\cal L}$ and ${\cal L}_x$ are
completely equivalent. In particular, the conjugate momenta of $X^M$ are
related to each other by exchange of $F_i^m$ and $\pa_iX^m$
\be
 \frac{\pa{\cal L}_x}{\pa (\pa_i X^M)}=\left.\frac{\pa{\cal L}}
 {\pa (\pa_iX^M)}\right|_{\pa X=F} \label{st_momrel}
\ee
with their internal and external components explicitly given by
\bea
 \left.\frac{\pa{\cal L}_x}{\pa (\pa_iX^m)}\right|_{F=\pa X} &=&
     \sg\g^{ij}\pa_jX^n\gh_{mn}+\e^{ij}\pa_jX^n\bh_{mn}+ 
     \sg\g^{ij}\pa_jX^\n \gh_{m\n}+\e^{ij}\pa_jX^\n\bh_{m\n} \label{sti}\\
 \left.\frac{\pa{\cal L}_x}{\pa (\pa_iX^\m)}\right|_{F=\pa X} &=&
     \sg\g^{ij}\pa_jX^n\gh_{\m n}+\e^{ij}\pa_jX^n\bh_{\m n}+ 
     \sg\g^{ij}\pa_jX^\n \gh_{\m\n}+\e^{ij}\pa_jX^\n \bh_{\m\n}\; .
 \label{ste}
\eea
Similarly, the energy momentum tensor $T^{(x)}_{ij}$ of ${\cal L}_x$
can be expressed as
\be
 T^{(x)}_{ij}\equiv\frac{1}{\sg}\frac{\pa{\cal L}_x}{\pa\g^{ij}}
             =\frac{1}{\sg}\left.\frac{\pa{\cal L}}{\pa\g^{ij}}
              \right|_{\pa X=F}\equiv \left.T_{ij}\right|_{\pa X=F}\; .
     \label{st_Trel}
\ee
We now observe that the equation of motion and the Bianchi identity for
$X^m$ can be written in the following form
\be
 \pa_i\left(\ba{c} \frac{\pa{\cal L}}{\pa (\pa_iX^m)}\\
                   \e^{ij}\pa_jX^m\ea\right) = 0
 \label{st_eomb}
\ee
where the conjugate momenta are given by eq.~\refs{sti}.
This suggests the existence of a duality symmetry rotating conjugate momenta
and the dual field strengths $\e^{ij}\pa_jX^m$  of the internal target space
coordinates into each other. Such a symmetry should leave the other equations
of motion (the ones for $X^\m$) invariant. Later, we will see that
already the external conjugate momenta~\refs{ste} are invariant by themselves.

\vspace{0.4cm}

Equation~\refs{st_eomb} shows an apparent asymmetry between its upper
and lower part and is clearly not written in a manifest duality invariant
form. To find such a form it is useful to introduce a dual Lagrangian
${\cal L}_y$ with the r\^ole of equations of motion and Bianchi identities
being exchanged. We dualize the internal coordinates $X^m$ to coordinates
$Y_m$ by defining
\be
 {\cal L}_y = \left.{\cal L}\right|_{\pa X=F}+\e^{ij}\pa_iY_mF_j^m\; .
 \label{st_Ly}
\ee
Then the conjugate momentum for $Y_m$ is given by
\be
 \frac{\pa{\cal L}_y}{\pa (\pa_iY_m)} = \e^{ij}F_j^m\; .\label{st_ymom}
\ee
This leads to the equation of motion $\e^{ij}\pa_iF_j^m=0$ which implies
that $F_i^m=\pa_iX^m$ locally. Furthermore, from the definition of
${\cal L}_y$ and the eqs.~\refs{st_LLx}, \refs{st_momrel}, \refs{st_Trel} we
have the following relations
\bea
 \frac{\pa{\cal L}_y}{\pa F_i^m} &=& \frac{\pa{\cal L}_x}{\pa \pa_i X^m} 
                                     -\e^{ij}\pa_jY_m  = 0 \label{ste1}\\
 \frac{\pa{\cal L}_y}{\pa (\pa_iX^\m )} &=&
                  \frac{\pa{\cal L}_x}{\pa (\pa_iX^\m )} \label{ste2}\\
 \frac{\pa{\cal L}_y}{\pa X^\m } &=&
                  \frac{\pa{\cal L}_x}{\pa X^\m }\label{ste3} \\
 T^{(y)}_{ij}\equiv\frac{1}{\sg}\frac{\pa{\cal L}_y}{\pa\g^{ij}}&=&T^{(x)}_{ij}
 \label{ste4}
\eea
Taking the derivative $\pa_i$ of the first of these equations shows that
the ${\cal L}_y$ equation of motion for $F_i^m$ implies the ${\cal L}_x$
equation of motion for $X^m$. The second and third equation show that
${\cal L}_y$ and ${\cal L}_x$ lead to the same equations of motion for
$X^\m$ and, finally, the fourth equation shows the equality of the
energy momentum tensors. Therefore, the theories defined by
${\cal L}_x$ and ${\cal L}_y$ are classically equivalent.

\vspace{0.4cm}

We are now able to pair the internal conjugate momenta and the dual
field strengths in a more symmetric way. Using the eqs.~\refs{ste1} and
\refs{st_ymom} along with the internal conjugate momentum of ${\cal L}_x$
in eq.~\refs{sti} and $F_i^m=\pa_iX^m$ we find
\bea
 \e^{ij}\pa_jY_m &=& \frac{\pa{\cal L}_x}{\pa (\pa_i X^m)} = 
                     \gh_{mn}\sg\g^{ij}\pa_jX^n+\bh_{mn}\e^{ij}\pa_jX^n+ 
                      \gh_{m\n}\sg\g^{ij}\pa_jX^\n \nn \\
                 &&\quad\quad\quad\quad\quad + \bh_{m\n}\e^{ij}\pa_jX^\n
 \label{st_up} \\
 \e^{ij}\pa_jX^m &=& \frac{\pa{\cal L}_y}{\pa (\pa_iY_m)} \label{st_down}
\eea
This shows that the internal conjugate momentum of ${\cal L}_x$ equals the
dual field strength of ${\cal L}_y$ and vice versa. As a final step we would
like to find an explicit expression for the ${\cal L}_y$ conjugate
momentum which is still missing on the right hand side of the second equation
above. This can be done by solving for $F_i^m=\pa_iX^m$ in terms of $Y_m$
by using the equation of motion~\refs{ste1} for $F_i^m$. Explicitly, this
equation reads
\be
 \sg\g^{ij}F_j^n\gh_{mn}+\e^{ij}F_j^n\bh_{mn}+ 
  \sg\g^{ij}\pa_jX^\n \gh_{m\n}+\e^{ij}\pa_jX^\n \bh_{m\n}-\e^{ij}\pa_jY_m
    = 0\; .
  \label{st_solv}
\ee
Its solution is given by
\be 
 F_i^m = \ght^{mn}\frac{1}{\sg}{\e_i }^j\pa_jY_n+\bht^{mn}\pa_iY_n
         +{\ght^m }_\n\frac{1}{\sg}{\e_i }^j\pa_jX^\n
         +{\bht^m }_\n\pa_iX^\n\; ,
 \label{st_fsol}
\ee
where the dual background fields are defined by
\bea
 \ght_{mn} &=& \gh_{mn}+\bh_{mr}\gh^{rs}\bh_{ns} \nn \\
 \bht^{mn} &=& -\ght^{mr}\bh_{rs}\gh^{sn} \nn \\
 {\ght^m }_\n &=& -\ght^{mr}\bh_{r\n}-\bht^{mr}\gh_{r\n} \label{st_dualb}\\
 {\bht^m }_\n &=& -\bht^{mr}\bh_{r\n}-\ght^{mr}\gh_{r\n} \nn \; .
\eea
Inserting this result into eq.~\refs{st_down} leads to the explicit expression
for the conjugate momentum of $Y_m$. Then~\refs{st_up} and
\refs{st_down} take the form
\bea
  \e^{ij}\pa_jY_m = \frac{\pa{\cal L}_x}{\pa \pa_i X^m} &=& 
                     \gh_{mn}\sg\g^{ij}\pa_jX^n+\bh_{mn}\e^{ij}\pa_jX^n\nn\\
                    &&+\gh_{m\n}\sg\g^{ij}\pa_jX^\n+ \bh_{m\n}\e^{ij}\pa_jX^\n
  \label{st_up1} \\
  \e^{ij}\pa_jX^m = \frac{\pa{\cal L}_y}{\pa (\pa_iY_m)} &=&
                    \ght^{mn}\sg\g^{ij}\pa_jY_n+\bht^{mn}\e^{ij}\pa_jY_n\nn\\ 
                    &&+{\ght^m }_\n\sg\g^{ij}\pa_jX^\n+
                    {\bht^m }_\n\e^{ij}\pa_jX^\n\; .
  \label{st_down1}
\eea
These two equations are identical in form and can be obtained from each
other by an exchange of $X^m$ with $Y_m$ and the background fields with
their duals as defined in eq.~\refs{st_dualb}. This result could be expected,
and reflects the well known fact that the dual Lagrangian ${\cal L}_y$
expressed in terms of $Y_m$ (computed from eq.~\refs{st_Ly} by integrating
out $F_i^m$ via eq.~\refs{st_fsol}) is of the same form as ${\cal L}$.
As we will see this is no longer true in the case of the membrane.

In order to read off the symmetry from eqs.~\refs{st_up1} and
\refs{st_down1} we solve for the vertex operators $\e^{ij}\pa_jY_m$
and $\e^{ij}\pa_jX^m$ which still appear on the right hand side.
This leads to
\bea
 \left(\ba{l} \e^{ij}\pa_jY_m \\ \e^{ij}\pa_jX^m\ea\right) &=&
 \left(\ba{cc} \ght_{mn} & \bh_{mr}\gh^{nr} \\
               \gh^{mr}\bh_{nr} & \gh^{mn} \ea\right)
 \left[ \left(\ba{l} \sg\g^{ij}\pa_jX^n\\ \sg\g^{ij}\pa_jY_n
  \ea\right) +\left(\ba{l} \gh^{nr}\gh_{r\n} \\ \ght_{nr}{\ght^r }_\n
  \ea\right)\sg\g^{ij}\pa_jX^\n \nn \right. \\
 &&\left. + \left(\ba{l} \gh^{nr}\bh_{r\n} \\
    \ght_{nr}{\bht^r }_\n\ea\right)\e^{ij}\pa_jX^\n\right]\; . \label{st_dubl}
\eea
Let us introduce the following abbreviations
\bea
 G_{mn} &=& \gh_{mn} \nn \\
 B_{mn} &=& \bh_{mn} \nn \\
 A_\m^{(1)n} &=& \gh^{nr}\gh_{\m r} \\
 A_{\m n}^{(2)} &=& \ght_{nr}{\ght^r }_\m = \bh_{\m n}+B_{nr}A_\m^{(1)r} \nn \\
 V_\m^{(1)n} &=& \gh^{nr}\bh_{r\m} \nn \\
 V_{\m n}^{(2)} &=& \ght_{nr}{\bht^r }_\m = -\gh_{\m n}+B_{nr}V_\m^{(1)r} \nn
\eea
Furthermore, to write eq.~\refs{st_dubl} in a more compact form we introduce
quantities in the $2d$--dimensional vector space spanned by the internal
coordinates $X^m$ and their duals $Y_m$. First we define 
the $2d\times 2d$ matrix (using matrix notation for $G_{mn}$ and $B_{mn}$)
\be
 M = \left(\ba{cc} G-B^TG^{-1}B & BG^{-1} \\ -G^{-1}B & G^{-1}\ea\right)\; ,
 \label{st_M}
\ee
which contains the metric and antisymmetric tensor moduli and serves as
a parameterization of the moduli space. The two vectors
\be
 {\bf A}_\n = \left(\ba{l} A_\n^{(1)n} \\ A_{\n n}^{(2)}\ea\right)\; ,\quad
 {\bf V}_\n = \left(\ba{l} V_\n^{(1)n} \\ V_{\n n}^{(2)}\ea\right)
 \label{st_vec}
\ee
contain the vector fields of the theory; that is, the graviphotons and the
ones arising form the antisymmetric tensor. Finally, we use the
following short hand notation for the vertex operators
\be
 \wtd{\cF} = \left(\ba{l}  \e^{ij}\pa_jX^m\\ \e^{ij}\pa_jY_m\ea\right)
   \; ,\quad
 \cF = \left(\ba{l} \sg\g^{ij}\pa_jX^n \\ \sg\g^{ij}\pa_jY_n\ea\right)
\ee
\be
 {\cal A}^\n = \sg\g^{ij}\pa_jX^\n\; , \quad
 {\cal V}^\n = \e^{ij}\pa_jX^\n \; .
\ee
Note, that $\cF$ and $\wtd{\cF}$ are vectors on the $(X^m,Y_m)$
space whereas ${\cal A}^\n$ and ${\cal V}^\n$ are scalars. Since the
worldvolume index $i$ appears as an overall index in what follows it has
been suppressed in this notation. With these definitions,
eq.~\refs{st_down1} can be written as
\be
 \eta\wtd{\cF} = M\left( \cF+{\bf A}_\n{\cal A}^\n+
                  {\bf V}_\n{\cal V}^\n\right) \label{st_dubl1}
\ee
with $\eta$ given by
\[
 \eta = \left(\ba{ll} 0 & {\bf 1}_d \\ {\bf 1}_d & 0\ea\right)\; .
\]
Using this form of the internal equations of motion, we are now in a position
to discuss duality rotations. We start by performing the following
transformation on the vertex operators
\bea
 \wtd{\cF} \rightarrow {P^{-1}}^T\wtd{\cF}\; ,&&
 \cF \rightarrow {P^{-1}}^T\cF \nn \\
 {\cal A}^\n \rightarrow {\cal A}^\n \; ,&&
 {\cal V}^\n \rightarrow {\cal V}^\n \label{st_vtrans}
\eea
where $P$ is an invertible $2d\times 2d$ matrix. Note, that we have not
transformed the external vertex operators ${\cal A}^\n$ and ${\cal V}^\n$,
as suggested by their index structure. In order to keep the form of
eq.~\refs{st_dubl1} invariant under this transformation we should
counter rotate the background as
\be
 M \rightarrow PMP^T \; ,\quad
 {\bf A}_\n \rightarrow {P^{-1}}^T{\bf A}_\n\; ,\quad 
 {\bf V}_\n \rightarrow {P^{-1}}^T{\bf V}_\n \; .
 \label{st_btrans}
\ee
In general, the transformation law for $M$ will not preserve
its structure so that we have to restrict the set of allowed matrices $P$.
We notice that, from its definition~\refs{st_M}, $M$ is symmetric and
fulfills the equation $M\eta M =\eta$. The latter property means that $M$
is an element of $O(d,d)$. In fact, $M$ is a parameterization of the moduli
coset $O(d,d)/O(d)\times O(d)$. Therefore, the group of duality rotations
which leaves eq.~\refs{st_dubl1} invariant is given by $O(d,d)$ so that $P$
is constrained by
\be
 P^T\eta P = \eta\; \label{st_P}.
\ee
The quantities ${\bf A}_\n$ and ${\bf V}_\n$ contain the same
degrees of freedom, namely the $2d$ vector fields of the reduced theory.
One might therefore ask, whether the transformations assigned to them in
eq.~\refs{st_btrans} are compatible with each other. From the definitions
of ${\bf A}_\n$ and ${\bf V}_\n$, eq.~\refs{st_vec}, and the definition of $M$,
eq.~\refs{st_M}, we deduce
\be
 {\bf V}_\n = -\eta M{\bf A}_\n \label{st_vcons}
\ee
which is indeed consistent with the transformations~\refs{st_btrans}
using eq.~\refs{st_P}.

\vspace{0.4cm}

So far, we have assigned $O(d,d)$ transformation properties to the moduli
in $M$ and the vector fields in ${\bf A}_\n$ and ${\bf V}_\n$. Clearly,
we would also like obtain the $O(d,d)$ properties of the external
metric $\gh_{\m\n}$ and the external antisymmetric tensor $\bh_{\m\n}$.
They can be read off from the external equations of motion for $X^\m$,
once these are brought into a manifest $O(d,d)$ invariant form. In conformal
gauge, this has been done in ref.~\cite{ms}. Here, we will not present the
invariant form of the full equations of motion but concentrate on the conjugate
momenta for $X^\m$. Using the expression~\refs{ste} for these conjugate
momenta, and inserting the eqs.~\refs{st_dubl} repeatedly, we arrive at
\be
 \frac{\pa{\cal L}_x}{\pa (\pa_iX^\m)} = 
  \bar{g}_{\m\n}{\cal A}^{i\n} + \left( B_{\m\n}
  +\frac{1}{2}{\bf A}_\m^T\eta{\bf A}_\n\right){\cal V}^{i\n} +
  {\bf A}_\m^T\eta\wtd{\cF}^i
 \label{st_sing}
\ee
with
\bea
 \bar{g}_{\m\n} &=& \gh_{\m\n}-A_\m^{(1)n}A_{\n n}^{(1)} \label{st_ginv}\\
 B_{\m\n} &=& \bh_{\m\n}+\frac{1}{2}A_\m^{(1)r}A_{\n r}^{(2)}
              - \frac{1}{2}A_\n^{(1)r}A_{\m r}^{(2)}
              -B_{rs}A_\m^{(1)r}A_\n^{(1)s}\; .\label{st_binv}
\eea
Given the known transformations~\refs{st_btrans}, we conclude that the
external conjugate momenta are invariant if $\bar{g}_{\m\n}$ and
$B_{\m\n}$ as defined above are $O(d,d)$ singlets, that is
\be
 \bar{g}_{\m\n}\rightarrow \bar{g}_{\m\n}\; ,\quad
 B_{\m\n}\rightarrow B_{\m\n}\; . \label{st_btrans1}
\ee

\vspace{0.4cm}

To summarize, for a $\DH$--dimensional string $\s$--model dimensionally
reduced to $D=\DH -d$ dimensions, we have demonstrated the existence of
an $O(d,d)$ symmetry which leaves the equations of motion invariant.
This symmetry acts on the vertex operators as in eq.~\refs{st_vtrans}
and on the background fields as in eq.~\refs{st_btrans},
\refs{st_btrans1}. An independent check for the background transformation
laws is provided by the effective low energy action. Its dimensional
reduction to $D$ dimensions should lead to a theory which is invariant under
the $O(d,d)$ symmetry, acting on the background fields. In particular, the
reduced effective action should be expressible in terms of invariant
combinations of the $O(d,d)$ covariant quantities $M$,
${\bf A}_\n$, $\bar{g}_{\m\n}$ and $B_{\m\n}$. That this is indeed true
has been shown in ref.~\cite{ms}. For the membrane, we will use this
observation to confirm the background transformations obtained from
the worldvolume theory by dimensional reduction of $\DH =11$
supergravity.

\section{Duality on the Membrane Worldvolume}

Now we would like to apply the method of the previous section to the
bosonic part of the $\DH =11$ supermembrane~\cite{hlp,kappa,duff_rep}.
For the moduli part this has first been done by Duff and Lu~\cite{dl}.
Here we will keep the full background field content of the theory.

We denote the three worldvolume coordinates of the membrane by $\x^i$,
$i,j,k,...=0,1,2$ and the worldvolume metric with signature $(-++)$ by
$\g_{ij}$. The target space coordinates $X^M(\x^i )$ are indexed
by uppercase letter $M,N,P,...=0,...,10$. The bosonic background field
content of the supermembrane is given by a metric $\gh_{MN}=\gh_{MN}(X^R )$
and a 3--form field $\bh_{MNP}=\bh_{MNP}(X^R )$. With these definitions,
the bosonic part of the supermembrane Lagrangian reads
\be
 {\cal L} = \frac{1}{2}\sg\g^{ij}\pa_iX^M\pa_jX^N\gh_{MN}
            +\frac{1}{6}\e^{ijk}\pa_iX^M\pa_jX^N\pa_kX^P\bh_{MNP}
            -\frac{1}{2}\sg\; .
 \label{membr}
\ee
Note that this Lagrangian contains a cosmological constant term. For the
string $\s$-model such a term was forbidden by conformal invariance.
This leads to a modified expression for the energy momentum tensor
\be
 T_{ij}\equiv\frac{1}{\sg}\frac{\pa{\cal L}}{\pa\g^{ij}}
       =\frac{1}{2}\left( \pa_iX^M\pa_jX^N\gh_{MN}-\g_{ij}\g^{kl}
        \pa_kX^M\pa_lX^N\gh_{MN}+\frac{1}{2}\g_{ij}\right) = 0\; .
        \label{m_T}
\ee
The vanishing of $T_{ij}$ now implies that $\g_{ij}$ is the induced metric
on the worldvolume
\be
 \g_{ij} = \pa_iX^M\pa_jX^N\gh_{MN}\; .\label{indmetric}
\ee
For the dimensional reduction of the Lagrangian~\refs{membr} to
$D=\DH -d$ dimensions we split the target space coordinates as
$X^M =(X^\m ,X^m )$. The external coordinates $X^\m$ are indexed by
$\m ,\n ,\r ,...= 0,...,D-1$ and the internal coordinates $X^m$ by
$m,n,r,...=D,...,\DH -1$. We assume that the background fields are
independent on the internal coordinates $X^m$; that is, 
$\gh_{MN} = \gh_{MN}(X^\m )$ and $\bh_{MNR} = \bh_{MNR}(X^\m )$.
Under this split of coordinates, the background fields break up as
$\gh_{MN}=(\gh_{\m\n},\gh_{\m n},\gh_{mn})$,
$\bh_{MNR}=(\bh_{\m\n\r},\bh_{\m\n r},\bh_{\m nr},\bh_{mnr})$.
The background field content of the reduced theory is therefore given
by the metric $\gh_{\m\n}$ and the 3--form $\bh_{\m\n\r}$, the $d$
2--forms $\bh_{\m\n r}$, the $d(d+1)/2$ vector fields $\gh_{\m n}$,
$\bh_{\m nr}$ and the $d(d^2+5)/6$ moduli fields $\gh_{mn}$, $\bh_{mnr}$.
The Lagrangian~\refs{membr} can be decomposed as
\be
 {\cal L} = \sum_{n=0}^3{\cal L}^{(n)}
\ee
with ${\cal L}^{(n)}$ being homogeneous of degree $n$ in the internal
coordinates $X^m$. Explicitly, these various parts are given by
\bea
 {\cal L}^{(0)} &=& \frac{1}{2}\sg\g^{ij}\pa_iX^\m\pa_jX^\n \gh_{\m\n}
            +\frac{1}{6}\e^{ijk}\pa_iX^\m\pa_jX^\n\pa_kX^\r\bh_{\m\n\r}
            -\frac{1}{2}\sg\nn \\
 {\cal L}^{(1)} &=& \sg\g^{ij}\pa_iX^\m\pa_jX^n \gh_{\m n}
            +\frac{1}{2}\e^{ijk}\pa_iX^\m\pa_jX^\n\pa_kX^r \bh_{\m\n r}\nn \\
 {\cal L}^{(2)} &=& \frac{1}{2}\sg\g^{ij}\pa_iX^m\pa_jX^n \gh_{mn}
            +\frac{1}{2}\e^{ijk}\pa_iX^\m\pa_jX^n\pa_kX^r \bh_{\m nr}
 \label{m_split} \\
 {\cal L}^{(3)} &=& \frac{1}{6}\e^{ijk}\pa_iX^m\pa_jX^n\pa_kX^r \bh_{mnr}
  \nn\; .
\eea

Since ${\cal L}$ does not depend on $X^m$ explicitly, we may
introduce first order fields $F_i^m$ and rewrite the Lagrangian as
\be
 {\cal L}_x = \sum_{n=0}^3(1-n)\left. {\cal L}^{(n)}\right|_{\pa X=F}
              +\left. \frac{\pa{\cal L}}
              {\pa (\pa_i X^m)}\right|_{\pa X=F}\pa_i X^m\; ,
 \label{m_Lx}
\ee
where $|_{\pa X =F}$ indicates that $\pa_i X^m$ has been replaced by
$F_i^m$ in the respective expression. As in the string case, we can prove
the equivalence of ${\cal L}$ and ${\cal L}_x$ by analyzing the equation
of motion
\be
 \frac{\pa {\cal L}_x}{\pa F_i^m} = 0
\ee
for $F_i^m$. Its solution, $F_i^m=\pa_iX^m$, substituted into ${\cal L}_x$
results in
\be
 {\cal L} = \left. {\cal L}_x\right|_{F=\pa X}\; . \label{m_LLx}
\ee 
In particular, we have the following relations between the conjugate
momenta of ${\cal L}$ and ${\cal L}_x$
\be
 \frac{\pa{\cal L}_x}{\pa (\pa_i X^M)}=\left.\frac{\pa{\cal L}}
 {\pa (\pa_iX^M)}\right|_{\pa X=F} \label{m_momrel}
\ee
and the energy momentum tensors of ${\cal L}$ and ${\cal L}_x$
\be
 T^{(x)}_{ij}\equiv\frac{1}{\sg}\frac{\pa{\cal L}_x}{\pa\g^{ij}}
             =\frac{1}{\sg}\left.\frac{\pa{\cal L}}{\pa\g^{ij}}
              \right|_{\pa X=F}\equiv\left.T_{ij}\right|_{\pa X=F}\; .
              \label{m_Trel}
\ee
The latter relation implies that $\g_{ij}$ is still given by the induced
metric as in eq.~\refs{indmetric}, but with $\pa_iX^m$ replaced by $F_i^m$.
That is,
\be
 \g_{ij} = F_i^mF_j^n\gh_{mn}+F_i^m\pa_jX^\n\gh_{m\n}+\pa_iX^\m F_j^n\gh_{\m n}
           +\pa_iX^\m\pa_jX^\n\gh_{\m\n}\; .
 \label{indmetric1}
\ee
We remark that in ref.~\cite{dl} only the first term in eq.~\refs{indmetric1}
was taken into account. However, even if all background fields except the
moduli are turned off, the last term in eq.~\refs{indmetric1} is still
nonvanishing. As we will see, the presence of these extra terms
complicates the subsequent calculation considerably.

The explicit expressions for the conjugate momenta are given by
\bea
 \left.\frac{\pa{\cal L}_x}{\pa (\pa_iX^m)}\right|_{F=\pa X} &=&
  \sg\g^{ij}\pa_jX^N\gh_{mN}+\frac{1}{2}\e^{ijk}\pa_jX^N\pa_kX^R\bh_{mNR} 
  \label{mi} \\
 \left.\frac{\pa{\cal L}_x}{\pa (\pa_iX^\m)}\right|_{F=\pa X} &=&
  \sg\g^{ij}\pa_jX^N\gh_{\m N}+\frac{1}{2}\e^{ijk}\pa_jX^N\pa_kX^R
    \bh_{\m NR}\; . \label{me}
\eea
As for the string, we would now like to pair the equations of motion
for $X^m$ with Bianchi identities. What are the vertex operators for
these Bianchi identities in the case of the membrane? A natural
generalization of the string winding operator is $\e^{ijk}\pa_jX^m\pa_kX^n$,
which leads to the conservation equation
\be
 \pa_i\left(\ba{c} \frac{\pa{\cal L}}{\pa (\pa_iX^m)}\\
                   \e^{ijk}\pa_jX^m\pa_kX^n\ea\right) = 0
 \label{m_eomb}
\ee
with the conjugate momentum given in eq.~\refs{mi}.
Note that we have paired $d$ conserved momenta and $d(d-1)/2$ conserved
winding numbers in the above equation. Unlike for the string, these
numbers are not equal except for a dimensional reduction to $D=8$ ($d=3$).
This reflects the fact that there exist $d(d-1)/2$ ways for the
two spatial directions of the membrane to wrap around $d$ compact
directions. That the pairing in eq.~\refs{m_eomb} is reasonable is also
suggested by the result in ref.~\cite{ss}. There it has been shown that
an analogous pairing for the $5$ brane leads to charges which transform
correctly under the $SL(2)$ S--duality of string theory.
Another important difference from the string case is the existence of
an additional ``mixed'' Bianchi identities
\be
 \pa_i\left(\e^{ijk}\pa_jX^m\pa_kX^\n\right) = 0\; . \label{m_mixb}
\ee
We interpret these conserved currents as corresponding to membrane
states with only one spatial coordinate wrapped around a compact direction.

Again, we would like to find a symmetric form of eq.~\refs{m_eomb}
which shows the inferred duality symmetry in a manifest way. To do this,
we construct a dual Lagrangian with the r\^ole of equations of motion
and Bianchi identities exchanged. Having noticed the presence of
the two types of Bianchi identities in eq.~\refs{m_eomb} and \refs{m_mixb},
a natural definition for this dual Lagrangian ${\cal L}_y$ is
\be
 {\cal L}_y = \left.{\cal L}\right|_{\pa X=F}+\e^{ijk}\pa_iY_{mn}
              F_j^mF_k^n + \e^{ijk}\pa_iY_{m\n}F_j^m\pa_kX^\n \; .
 \label{m_Ly}
\ee
We have introduced two types of coordinates ``dual'' to $X^m$, namely
$Y_{mn}$ and $Y_{m\n}$. Therefore ${\cal L}_y$, with
the auxiliary field $F_i^m$ being integrated out, will be different in
structure from ${\cal L}_x$. In particular, the target space dimension of
${\cal L}_y$ does not coincide with the one of ${\cal L}_x$. This is to be
contrasted to the string case where both Lagrangians were of the same form.

To establish a relation between ${\cal L}_y$ and ${\cal L}_x$
we consider the conjugate momenta
\bea
 \frac{\pa{\cal L}_y}{\pa (\pa_iY_{mn})} &=& \e^{ijk}F_j^mF_k^n
     \label{m_ymom} \\
 \frac{\pa{\cal L}_y}{\pa (\pa_iY_{m\n})} &=& \e^{ijk}F_j^m\pa_kX^\n
     \label{m_ymomm}\; .
\eea
and the corresponding equations of motion
\bea
 \pa_i\left(\e^{ijk}F_j^mF_k^n\right) &=& 0 \label{m_eomy}\\
 \pa_i\left(\e^{ijk}F_j^m\pa_kX^\n\right) &=& 0\; . \label{m_eomym}
\eea 
Certainly, these equations are solved by $F_i^m=\pa_iX^m$. Is this really
the only solution of the system~\refs{m_eomy}, \refs{m_eomym}? Let us
consider a specific example for $d=2$. We take $F_i^1 = (\x^0\x^1,0,0)$
and $F_i^2=(0,1,0)$ which fulfill $\pa_i(\e^{ijk}F_j^1F_k^2)=0$; that is,
the $Y_{mn}$ equation of motion~\refs{m_eomy}. On the other hand
$\e^{ijk}\pa_jF_k^1 = (0,0,-\x^0)$, which implies that $F_i^1$ is not closed
(as a 1--form on the world volume). This shows that adding the first
Lagrange multiplier term in eq.~\refs{m_Ly} only, as was done in
refs.~\cite{dl}, does not guarantee $F_i^m=\pa_iX^m$. Since eq.~\refs{m_eomy}
provides $d(d-1)/2$ linear homogeneous equations for the $3d$ quantities
$\e^{ijk}\pa_jF_k^m$, we expect similar examples up to $d=6$ at least.
Certainly, the additional mixed Bianchi identities~\refs{m_eomym} eliminate
some of these cases. For the above example in $d=2$ we have
$\pa_i(\e^{ijk}F_j^1\pa_kX^\n )=-\x^0\pa_2X^\n$, so that it does not solve
the full system of equations~\refs{m_eomy}, \refs{m_eomym}
if only $\pa_2X^\n\ne 0$ for one $\n$. Such a condition, however, is not
guaranteed and can be violated for membranes oriented transversally
to the external space. Therefore, though being an ``improvement'' over
just using the first condition~\refs{m_eomy}, the system~\refs{m_eomy},
\refs{m_eomym} still does not force $F_i^m=\pa_iX^m$ in general.
A way to unambiguously obtain such a solution, is to replace the
two Lagrange multiplier terms in eq.~\refs{m_Ly} by $\e^{ijk}A_{jm}F_k^m$
where $A_{jm}$ are $d$ worldvolume vector fields. Their equation of motion
is given by $\e^{ijk}\pa_jF_k^m=0$ which implies $F_i^m=\pa_iX^m$ locally.
Such a method can, for example, be used to derive the type IIA 2D--brane
action from the 11--dimensional supermembrane~\cite{town}. In our context,
however, the worldvolume vectors $A_{im}$, unlike the fields
$Y_{mn}$, $Y_{m\n}$, are not the appropriate degrees of freedom to describe
winding modes of the membrane. Correspondingly, pairing the original
Lagrangian ${\cal L}_x$ with ${\cal L}_y$ defined in such a way does not
show any of the expected U--duality symmetry structure.
Alternatively, the term $\e^{ijk}A_{jm}F_k^m$ could be added to the 
definition~\refs{m_Ly} of ${\cal L}_y$. Though there is nothing wrong
with this in principle, it is hard to see what the interpretation of the
additional vector fields $A_{jm}$ could be.

For the moment, we will therefore accept the somewhat unfortunate situation
that the theory described by ${\cal L}_y$ in eq.~\refs{m_Ly} seems to be
more general than the original one and concentrate on those solutions of
${\cal L}_y$ for which $F_i^m=\pa_iX^m$. Then we have from eq.~\refs{m_Ly}
and the eqs.~\refs{m_LLx}, \refs{m_momrel}, \refs{m_Trel}
\bea
 \frac{\pa{\cal L}_y}{\pa F_i^m} &=& \frac{\pa{\cal L}_x}{\pa \pa_i X^m} 
                -2\e^{ijk}\pa_jY_{mn}F_i^n-2\e^{ijk}\pa_jY_{m\n}\pa_kX^\n
                                  = 0 \label{me1}\\
 \frac{\pa{\cal L}_y}{\pa (\pa_iX^\m )} &=&
                  \frac{\pa{\cal L}_x}{\pa (\pa_iX^\m )}
                  -2\e^{ijk}\pa_jY_{\m n}F_k^n \label{me2}\\
 \frac{\pa{\cal L}_y}{\pa X^\m } &=&
                  \frac{\pa{\cal L}_x}{\pa X^\m }\label{me3} \\
 T^{(y)}_{ij}\equiv\frac{1}{\sg}\frac{\pa{\cal L}_y}{\pa\g^{ij}}&=&
     T^{(x)}_{ij} \label{me4}\; .
\eea
The additional terms in eq.~\refs{me1} and \refs{me2} vanish once we set
$F_i^m=\pa_iX^m$ and take the derivative $\pa_i$. This shows the classical
equivalence of ${\cal L}_x$ and ${\cal L}_y$ provided the solutions of
${\cal L}_y$ are restricted to those with $F_i^m=\pa_iX^m$. From now on
we will assume this restriction and use $F_i^m$ and $\pa_iX^m$
interchangeably.

\vspace{0.4cm}

Let us now rewrite the internal conjugate momenta and the Bianchi identity
from eq.~\refs{m_eomb} in a more symmetric way using
the Lagrangian ${\cal L}_y$. Putting together eq.~\refs{me1},
the internal conjugate momentum~\refs{mi}, eq.~\refs{m_ymom} and
$F_i^m=\pa_iX^m$ we find
\bea
 2\e^{ijk}\pa_jY_{mN}\pa_kX^N = \frac{\pa{\cal L}_x}{\pa (\pa_i X^m)} &=& 
                   \gh_{mn}\sg\g^{ij}\pa_jX^n+\frac{1}{2}\bh_{mnr}
                   \e^{ijk}\pa_jX^n\pa_kX^r+\gh_{m\n}\sg\g^{ij}\pa_jX^\n \nn\\
                     &&+\bh_{mn\n}\e^{ijk}\pa_jX^n\pa_kX^\n
                       +\frac{1}{2}\bh_{m\r\n}\e^{ijk}\pa_jX^\r\pa_kX^\n
  \label{m_up} \\
 \e^{ijk}\pa_jX^m\pa_kX^n &=& \frac{\pa{\cal L}_y}{\pa (\pa_iY_{mn})}\; .
  \label{m_down}
\eea
By construction of the dual Lagrangian ${\cal L}_y$, the conjugate momentum
of $Y_{mn}$ equals the ``Bianchi identity'' $\e^{ijk}\pa_jX^m\pa_kX^n$
of ${\cal L}_x$. The conjugate momentum of $X^m$, on the other hand,
equals the operator $2\e^{ijk}\pa_jY_{mN}\pa_kX^N$ which we interpret as the
Bianchi identity of ${\cal L}_y$. Its unconventional form in terms of $X^N$
results because we have used Lagrange multiplier terms in eq.\refs{m_Ly} which 
are bilinear in $\pa X$. Moreover, according to eq.~\refs{m_eomym}, the
mixed Bianchi identity equals the conjugate momentum of $Y_{m\n}$
\be
 \e^{ijk}\pa_jX^m\pa_kX^\n = \frac{\pa{\cal L}_y}{\pa (\pa_iY_{m\n})}\; .
 \label{m_single}
\ee
As in the string case, we now have to find explicit expressions for the
right hand sides of eq.~\refs{m_down} and \refs{m_single} in terms
of the dual coordinates $Y_{mN}$. These expressions should be obtained
by solving eq.~\refs{me1}, which explicitly reads
\bea
 \sg\g^{ij}F_j^n\gh_{mn}+\frac{1}{2}\e^{ijk}F_j^nF_k^r\bh_{mnr}
    +\sg\g^{ij}\pa_jX^\n\gh_{m\n}+\e^{ijk}F_j^n\pa_kX^\r\bh_{mn\r}&& \nn \\
 +\frac{1}{2}\e^{ijk}\pa_jX^\n\pa_kX^\r\bh_{m\n\r}
   -2\e^{ijk}\pa_jY_{mn}F_k^n - 2\e^{ijk}\pa_jY_{m\n}\pa_kX^\n &=& 0\; .
 \label{m_solv}
\eea
Unfortunately, this equation cannot be simply solved for $F_i^m$ in terms
of $Y_{mN}$ as in the string case because of the terms quadratic in $F_i^m$
and the appearance of $F_i^m$ in the second to last term (compare with
eq.~\refs{st_solv} for the string). The best we can do at this point is to
either solve eq.~\refs{m_solv} iteratively or to find an implicit solution.
Unlike in the string case, it is therefore hard to find a closed form for the
dual Lagrangian ${\cal L}_y$ with the auxiliary field $F_i^m$ being
integrated out. Consequently, our main focus is on the equations of motion
from which we attempt to read off the duality symmetry. To be able to do so, we
are clearly interested in a closed form of these equations. Therefore,
we will look for an implicit rather than an iterative solution of
eq.~\refs{m_solv}. For all background fields except the moduli turned off,
such a solution has been found in ref.~\cite{dl} assuming the relation
$\g_{ij}=F_i^mF_j^n\gh_{mn}$.  As noted earlier, however, this
relation is really incomplete and should be supplemented with the other
terms in eq.~\refs{indmetric1}. Then, even in the pure moduli case, an
additional term from the external space arises so that
$\g_{ij}=F_i^mF_j^n\gh_{mn}+\pa_iX^\m\pa_jX^\n\eta_{\m\n}$ where $\eta_{\m\n}$
is the Minkowski metric of the external space. Unfortunately, in the
presence of this extra term the solution of ref.~\cite{dl} no longer works.
How, then, can we proceed with the most general
background when even the pure moduli case poses such problems?
The only systematic way out of this difficulty which we could find is to 
rewrite eq.~\refs{m_solv} in a ``full index range'' form as
\be
 \e^{ijk}V_{jmN}\pa_kX^N = -\sg\g^{ij}\pa_jX^N\gh_{mN}\; ,
 \label{m_solv1}
\ee
where we have defined
\be
 V_{jmN} = \frac{1}{2}\bh_{mRN}\pa_jX^R-2\pa_jY_{mN}\; ,
 \label{m_Vdef}
\ee
and $F_i^m$ has been replaced by $\pa_iX^m$. This equivalent form is
better adapted to the structure of the $\g_{ij}$ equation,
$\g_{ij}=\pa_iX^M\pa_jX^N\gh_{MN}$, and has the simple solution
\be
 V_{imN} = -\frac{1}{4\sg}{\e_i }^{jk}\pa_jX^R\pa_kX^S\gh_{mNRS}
 \label{m_Vsol}
\ee
with
\be
 \gh_{MNRS}\equiv \gh_{MR}\ \gh_{NS}-\gh_{MS}\ \gh_{NR}\; .
 \label{m_g4}
\ee
The verification is straightforward by inserting~\refs{m_Vsol} into
eq.~\refs{m_solv1} and using the fact that $\g_{ij}$ is given by the induced
metric~\refs{indmetric}. Note that is was crucial to rewrite the
equation~\refs{m_solv1} in the ``full index range'' form~\refs{m_solv1}
to find this solution. Moreover, as can be seen from the
definition~\refs{m_Vdef}, the off-diagonal dual coordinates $Y_{m\n}$ fit
nicely into this scheme.

Though~\refs{m_Vsol} appears to be the ``natural'' solution of
eq.~\refs{m_solv1}, it is unfortunately not the most general one. Any
$Z_{jmN}$ with $\e^{ijk}Z_{jmN}\pa_kX^N=0$ can be added to $V_{jmN}$ in
\refs{m_Vsol}. Clearly, a restriction to the subset of solutions with
$Z_{jmN}=0$ can be a source of symmetry breaking (if the subset is
noninvariant). Nevertheless, we will proceed with the solution~\refs{m_Vsol}
keeping in mind that we are actually using a certain subclass of solutions
of the dual theory. Later on we will comment on the remaining freedom
parameterized by $Z_{jmN}$.

\vspace{0.4cm}

Let us first concentrate on the $(mN)=(mn)$ part of the solution~\refs{m_Vsol},
which explicitly reads
\be
 V_{imn} = -\frac{1}{4\sg}{\e_i }^{jk}\left(\pa_jX^r\pa_kX^s\gh_{mnrs}+
           2\pa_jX^r\pa_kX^\s\gh_{mnr\s}+\pa_jX^\r\pa_kX^\s\gh_{mn\r\s}
           \right)\; .
\ee
This equation can be easily solved for $\pa_jX^m\pa_kX^n$ by inverting
$\gh_{mnrs}$ to give
\be
 \e^{ijk}\pa_jX^m\pa_kX^n = -\sg\g^{ij}\gh^{mnrs}V_{jrs}-\frac{1}{2}
                   \e^{ijk}\pa_jX^r\pa_kX^\s\gh_{r\s uv}\gh^{mnuv}-
                   \frac{1}{4}\e^{ijk}\pa_jX^\r\pa_kX^\s\gh_{\r\s uv}\gh^{mnuv}
 \; . \label{m_sol1}
\ee
Inserting this expression into equation~\refs{m_solv} eliminates the term
bilinear in $F_i^m$, and we get the implicit solution
\bea
 F_i^m &=& 2\ght^{mn}\frac{1}{\sg}{\e_i }^{jk}\pa_jY_{nN}\pa_kX^N
         -4\bht^{\nb m}\pa_iY_\nb +{\ght^m}_\n\pa_iX^\n \nn \\
        && -{\bht^m }_{r\s}\frac{1}{\sg}{\e_i }^{jk}\pa_jX^r\pa_kX^\s
         -{\bht^m }_{\r\s}\frac{1}{2\sg}{\e_i }^{jk}\pa_jX^\r\pa_kX^\s\; .
 \label{m_Fsol}
\eea
Here, we have introduced the notation $\mb = [m_1m_2]$ for an antisymmetric
pair of internal indices, which turns out to be useful in the following.
We define the summation over two of these index pairs to include a factor
$1/2$, that is
\be
 v_\mb w^\mb\equiv \frac{1}{2}v_{m_1m_2}w^{m_1m_2}
\ee
for any pair of antisymmetric tensors $v_\mb$ and $w^\mb$. The ``dual''
quantities in the above solution~\refs{m_Fsol} are defined by
\bea
 \ght_{mn} &=& \gh_{mn}+\bh_{m\rb}\gh^{\rb\sb}\bh_{n\sb} \nn \\
 \bht^{\mb n} &=& \ght^{nr}\bh_{r\sb}\gh^{\mb\sb} \nn \\
 {\ght^m}_\n &=& -\ght^{mr}\gh_{r\n}-\bht^{\rb m}\bh_{\n\rb} 
  \label{m_dualb} \\
 {\bht^m }_{r\s} &=& \ght^{ms}\bh_{sr\s}-\bht^{\sb m}\gh_{\sb r\s} \nn\\
 {\bht^m }_{\r\s} &=& \ght^{mr}\bh_{\r\s r}-\bht^{\sb m}\gh_{\sb\r\s} \nn \; .
\eea
It is interesting to observe how close these relations are to their
string analogs~\refs{st_dualb}. Basically, some of the indices have just
been replaced by antisymmetric index pairs. 

These results allow us to find an expression for the Bianchi identity of
$\e^{ijk}\pa_jX^m\pa_kX^n$ in terms of the dual coordinates
by using eq.~\refs{m_sol1} with $F_i^m$ replaced via eq.~\refs{m_Fsol}.
Then, from eqs.~\refs{m_up}, \refs{m_down} the completed pair of
conjugate momenta and Bianchi identities reads
\bea
 2\e^{ijk}\pa_jY_{mN}\pa_kX^N = \frac{\pa{\cal L}_x}{\pa (\pa_i X^m)} &=& 
                   \gh_{mn}\sg\g^{ij}\pa_jX^n+\frac{1}{2}\bh_{mnr}
                   \e^{ijk}\pa_jX^n\pa_kX^r\nn \\
                   &&+\gh_{m\n}\sg\g^{ij}\pa_jX^\n
                     +\bh_{nr\s}\e^{ijk}\pa_jX^r\pa_kX^\s\nn \\
                   &&+\frac{1}{2}\bh_{m\r\s}\e^{ijk}\pa_jX^\r\pa_kX^\s
  \label{m_up1} \\
 \e^{ijk}\pa_jX^{m_1}\pa_kX^{m_2} = \frac{\pa{\cal L}_y} 
                                    {\pa (\pa_iY_{m_1m_2})} &=& 
                   \ght^{\mb\nb}4\sg\g^{ij}\pa_jY_\nb + 
                   \bht^{\mb n}2\e^{ijk}\pa_jY_{nN}\pa_kX^N\nn \\
                   && + {\ght^\mb }_\n\sg\g^{ij}\pa_jX^\n
                      +{\bht^\mb }_{r\s}\e^{ijk}\pa_jX^r\pa_kX^\s\nn \\
                   && +\frac{1}{2}{\bht^\mb }_{\r\s}
                      \e^{ijk}\pa_jX^\r\pa_kX^\s  \; .
  \label{m_down1}
\eea
with the ``double indexed'' dual quantities defined by
\bea
 \ght^{\mb\nb} &=& \gh^{\mb\nb}-\bht^{\mb r}\ght_{rs}\bht^{\nb s} \nn \\
 {\ght^\mb }_\n &=& -\bht^{\mb n}\gh_{n\n}+\ght^{\mb\nb}\bht_{\nb\n} \nn \\
 {\bht^\mb }_{n\n} &=& -\bht^{\mb r}\bht_{rn\n}-\ght^{\mb\rb}\gh_{\rb n\n} \\
 {\bht^\mb }_{\r\n} &=& -\bht^{\mb r}\bht_{r\r\n}-\ght^{\mb\rb}\gh_{\rb\r\n}
   \nn\; .
\label{m_dualb1}
\eea
As in the string case, we would like to solve for the Bianchi identities of
$\e^{ijk}\pa_jY_{mN}\pa_kX^N$ and $\e^{ijk}\pa_jX^{m_1}\pa_kX^{m_2}$
which still appear on the right hand sides of the eqs.~\refs{m_up1}
and \refs{m_down1}. This leads to
\bea
 \left(\ba{l} 2\e^{ijk}\pa_jY_{mN}\pa_kX^N \\ \e^{ijk}\pa_jX^{m_1}\pa_kX^{m_2}
 \ea\right) &=& \left(\ba{cc} \ght_{mn} & \bh_{m\rb}\gh^{\rb\nb} \\
                            \gh^{\mb\rb}\bh_{n\rb} & \gh^{\mb\nb}\ea\right)
              \left[ \left(\ba{l} \sg\g^{ij}\pa_jX^n \\ \sg\g^{ij}\pa_jY_\nb\ea
                     \right)\right. \nn \\
            && +\left(\ba{l} \gh^{nr}\gh_{r\n} \\ \ght_{\nb\rb}
                     {\ght^\rb }_\n\ea\right) \sg\g^{ij}\pa_jX^\n
               + \left(\ba{l} \gh^{ns}\bh_{sr\s} \\ \ght_{\nb\sb}
                 {\bht^\sb }_{r\s}\ea\right) \e^{ijk}\pa_jX^r\pa_kX^\s\nn \\
            &&\left. +\frac{1}{2}\left(\ba{l} \gh^{ns}\bh_{s\r\s} \\
               \ght_{\nb\sb}{\bht^\sb }_{\r\s}\ea\right)
                \e^{ijk}\pa_jX^\r\pa_kX^\s\right]
\label{m_dubl}
\eea
Turning off all background fields except the moduli (which makes the
last three terms on the right hand side vanish) leaves us with an
equation similar to the one found in ref.~\cite{dl}. In particular the
moduli matrix is identical to the one found there. This means that we
are able to arrive at the correct moduli transformations. An explicit
example for this will be discussed in the next section. Note, however,
that we have derived this result taking the effect of the external
space into account. We are, therefore, truly dealing with a membrane in a
$\DH =11$--dimensional target space.

To write eq.~\refs{m_dubl} in a more compact form, we abbreviate
\bea
 G_{mn} &=& \gh_{mn} \nn \\
 G_{\mb\nb} &=& \gh_{\mb\nb} \nn \\
 B_{m\nb} &=& \bh_{m\nb} \nn \\
 A_\n^{(1)n} &=& \gh^{nr}\gh_{r\n} \label{m_fields}\\
 A_{\n\nb}^{(2)} &=& \ght_{\nb\rb}{\ght^\rb }_\n =
                     \bh_{\n\nb}-\bh_{\nb r}A_\n^{(1)r} \nn \\
 V_{r\s}^{(1)n} &=& \gh^{ns}\bh_{sr\s} \nn \\
 V_{r\s\nb}^{(2)} &=& \ght_{\nb\sb}{\bht^\sb }_{r\s} =
                      -\gh_{\nb r\s}-\bh_{\nb s}V_{r\s}^{(1)s} \nn \\
 B_{\n\s}^{(1)n} &=& \gh^{ns}\bh_{s\n\s} \nn \\
 B_{\n\s\nb}^{(2)} &=& \ght_{\nb\sb}{\bht^\sb }_{\n\s} =
                       -\gh_{\nb\n\s}-\bh_{\nb s}B_{\n\s}^{(1)s} \nn \; .
\eea
In addition, we introduce quantities in the $d(d+1)/2$--dimensional
space spanned by $(X^m,Y_\mb )$. Let us first consider the background fields.
We define the matrix
\be
 M= \F^k\left(\ba{cc} G_{mn}+B_{m\rb}G^{\rb\sb}B_{n\sb} & B_{m\rb}G^{\rb\nb} \\
                      G^{\mb\rb}B_{n\rb} & G^{\mb\nb} \ea\right)\; ,
 \label{m_M}       
\ee
which contains the metric and antisymmetric tensor moduli.
The $D$--dimensional dilaton $\F$ is given by
\be
 \F = \mbox{det}(G_{mn})\; .
\ee
Its appearance in the above definition with the specific power
\be
 k=\frac{1}{D-2}
\ee
is purely formal at this point and will be motivated below. The graviphotons
and 1--forms from the antisymmetric tensor are grouped into the vectors
\be
 {\bf A}_\n = \left(\ba{l} A_\n^{(1)m} \\ A_{\n\mb}^{(2)}\ea\right)\; ,
 {\bf V}_{r\n} = \F^{-k}\left(\ba{l} V_{r\n}^{(1)m} \\ 
                  V_{r\n\mb}^{(2)}\ea\right)\; .
 \label{m_vec}
\ee
Finally, we introduce the vector
\be
 {\bf B}_{\n\s} = \left(\ba{l} B_{\n\s}^{(1)m} \\ B_{\n\s\mb}^{(2)}\ea\right)
 \label{m_twof}
\ee
which contains the 2--forms and the graviphotons. Turning to the
vertex operators, we define
\bea
 \wtd{\cF} &=& \left(\ba{l} \wtd{\cF}_m^{(1)} \\ \wtd{\cF}^{(2)\mb}
                \ea\right) = \left(\ba{l}2\e^{ijk}\pa_jY_{mN}\pa_kX^N \\
                       \e^{ijk}\pa_jX^{m_1}\pa_kX^{m_2}\ea\right) \nn\\
 \cF &=& \left(\ba{l} \cF^{(1)m} \\ \cF_\mb^{(2)} \ea\right) =
           \F^{-k}\left(\ba{l} \sg\g^{ij}\pa_jX^m \\
                               4\sg\g^{ij}\pa_jY_\mb \ea\right)
 \label{m_F}
\eea
and
\be
 {\cal A}^\n = \F^{-k}\sg\g^{ij}\pa_jX^\n\; ,\quad
 {\cal V}^{r\n} = \e^{ijk}\pa_jX^r\pa_kX^\n\; ,\quad
 {\cal B}^{\n\s} = \e^{ijk}\pa_jX^\n\pa_kX^\s\; .\label{m_AVB}
\ee
Then eq.~\refs{m_dubl} can be written as
\be
 \wtd{\cF} = M\left( \cF+{\bf A}_\n{\cal A}^\n+{\bf V}_{r\n}{\cal V}^{r\n}+
             \frac{1}{2}{\bf B}_{\n\s}{\cal B}^{\n\s}
             \right)\; .
 \label{m_dubl1}
\ee
This equation is very similar in structure to the corresponding string
equation~\refs{st_dubl1}. One may therefore expect that $M$ transforms
as a tensor, $\wtd{\cF}$, $\cF$, ${\bf A}$, ${\bf V}$, ${\bf B}$ transform
as vectors and ${\cal A}$, ${\cal V}$, ${\cal B}$ transform as singlets under
an appropriate representation of the U--duality group in analogy to the
string case. To what extent this expectation
is really true will be worked out in detail in section 5. At this point,
we will continue to examine the other equations of motion of the theory
and attempt to write them in terms of the ``covariant'' quantities introduced
above.

Using the mixed part $(mN)=(m\n )$ of the solution~\refs{m_Vsol} along
with eq.~\refs{m_sol1}, we find for the mixed Bianchi identity~\refs{m_single}
\bea
 \e^{ijk}\pa_jX^m\pa_kX^\n =\frac{\pa{\cal L}_y}{\pa (\pa_iY_{m\n})} &=&
         -\gh^{mn}\bar{g}^{\n\r}\left[ (\bh_{nr\r}-\bh_{nrs}{\gh^s }_\r )
         \sg\g^{ij}\pa_jX^r +\right. \nn \\
      &&\left.   {\gh^r }_\r 4\sg\g^{ij}\pa_jY_{nr}
         +(\bh_{n\s\r}-{\gh^r }_\r\bh_{n\s r})\sg\g^{ij}\pa_jX^\s\right.\nn\\
      &&\left. -4\sg\g^{ij}\pa_jY_{n\r}\right]
         -{\gh^m }_\r\e^{ijk}\pa_jX^\r\pa_kX^\n
\eea
with
\be
 \bar{g}_{\m\n} = \gh_{\m\n}-{\gh_\m }^r\gh_{r\n} = \gh_{\m\n}-A_\m^{(1)r}
                  A^{(1)}_{\n r}\; .
\ee
In the previously introduced short hand notation this equation reads
\bea
 {\cal V}^{m\n} &=& -\F^{2k}\gh^{mn}g^{\n\r}\left[A_{\r nr}^{(2)}\cF^{(1)r}+
                    A_\r^{(1)r}\cF^{(2)}_{nr}+ \left( B_{\r\s n}+\frac{1}{2}
                    A^{(1)r}_\r A_{\s nr}^{(2)}\right.\right.\nn \\
                &&\left.\left. +\frac{1}{2}A_\s^{(1)r}A_{\r nr}^{(2)}
                   \right){\cal A}^\s -{\cal J}_{n\r }\right] -A_\r^{(1)m}
                   {\cal B}^{\r\n}
 \label{m_single1}
\eea
with
\bea
 g_{\m\n} &=& \F^k\bar{g}_{\m\n} = \F^k\left(\gh_{\m\n}-A_\m^{(1)r}
              A_{\n r}^{(1)}\right) 
 \label{m_g} \\
 B_{\m\n m} &=& \bh_{\m\n m}+\frac{1}{2}A_\m^{(1)r}A_{\n rm}^{(2)}
                -\frac{1}{2}A_\n^{(1)r}A_{\m rm}^{(2)}-B_{mrs}A_\m^{(1)r}
                A_\n^{(1)s}
 \label{m_B}
\eea
and the vertex operator
\be
 {\cal J}_{n\r} = \F^{-k}4\sg\g^{ij}\pa_jY_{n\r}\; .\label{m_J}
\ee
Note, that $g_{\m\n}$ in eq.~\refs{m_g} is the metric which naturally
appears in the dimensional reduction of the effective theory after
the Weyl rescaling. The definition~\refs{m_B} of the 2--forms is
motivated by the string analog~\refs{st_binv}.
Finally, we should find an ``appropriate'' form of the external conjugate
momentum. Using its explicit form~\refs{me} and inserting eq.~\refs{m_dubl}
we arrive at
\bea
 \frac{\pa{\cal L}_x}{\pa (\pa_iX^\m )} &=&
    g_{\m\n}{\cal A}^{i\n}+\left( B_{\n\m r}+\frac{1}{2}A_\n^{(1)s}
    A_{\m rs}^{(2)}+\frac{1}{2}A_\m^{(1)s}A_{\n rs}^{(2)}\right)
    {\cal V}^{ir\n} \nn \\
    && +{\bf A}^T_\m{\wtd{\cF}}^i+
       \frac{1}{2}(\bh_{\m\n\r}-{\gh^r }_\m\bh_{\n\r r}){\cal B}^{i\n\r}\; .
 \label{m_sing}
\eea
We would like to motivate the choice of dilaton powers which
we have included into various definitions of background quantities and
vertex operators. Clearly, the structure of the membrane equations of motion
does not uniquely fix the dilaton powers, which should be chosen in order to
get properly transforming quantities. We have therefore used the
information from the low energy theory that the correct invariant metric
after dimensional reduction to $D$ dimensions is the Weyl rescaled
metric $g_{\m\n}$ in eq.~\refs{m_g} and that the vector fields ${\bf A}_\n$
in eq.~\refs{m_vec} should not be rescaled (see section 6 for details).
This, together with the observation that the vertex operators in
$\wtd{\cF}$ and ${\cal V}$ should not be rescaled in order to preserve
the conservation equations $\pa_i{\wtd{\cF}}^i=0$ and $\pa_i{\cal V}^i=0$
essentially fixes the other powers of $\F$. In particular, it forces us
to include a factor $\F^{1/(D-2)}$ in the definition of $M$, eq.~\refs{m_M}.
In the next section we will see that this specific power of $\F$ is crucial
to discover the correct duality symmetry group.
 
In summary, we have attempted to rewrite the combined equations of
motion for ${\cal L}_x$ and ${\cal L}_y$ in the most symmetric form,
following the string analogy as closely as possible. The result
is given by the eqs.~\refs{m_dubl1}, \refs{m_single1}, \refs{m_sing}
along with the definitions~\refs{m_fields}, \refs{m_M}, \refs{m_vec},
\refs{m_twof}, \refs{m_g}, \refs{m_B} for the background fields and
\refs{m_F}, \refs{m_AVB}, \refs{m_J} for the vertex operators.

\section{Membrane Rotations for Moduli Backgrounds}

In this section, we would like to present an overview of duality
symmetries from the membrane worldvolume point of view and their relation to
the known U--duality groups of the the low energy supergravity theories
in diverse dimensions $D\geq 6$. To keep the discussion as simple as
possible, we will, in this section, restrict ourselves to backgrounds with
all fields except the moduli turned off. This allows us to point out some of
the major problems in a simple setting and motivates why, in the next section,
we will concentrate on the case $D=8$ to examine the situation when
all background fields are present.

We begin with the observation that the full system of equations of motion
and Bianchi identities (for a general background)~\refs{m_dubl1},
\refs{m_single1}, \refs{m_sing} has a manifest $GL(d)$ symmetry acting on
the internal indices $m,n,...$. It is the global part of the internal
coordinate transformations and certainly a subgroup of the full
duality group $\bG$. Let us now attempt to read off this duality
group from the moduli part of eq.~\refs{m_dubl1}; that is,
\be
 \wtd{\cF} = M\cF \label{m_mdub}
\ee
with $\wtd{\cF}$, $\cF$ and $M$ as defined in eq.~\refs{m_F} and \refs{m_M}.
Recall that $\wtd{\cF}$, $\cF$ are $d(d+1)/2$--dimensional vectors
containing vertex operators with the first $d$ entries corresponding to
conjugate momenta and the other $d(d-1)/2$ entries corresponding to
Bianchi identities of the original membrane worldvolume theory. The matrix
$M$ contains the $d(d^2+5)/6$ moduli $G_{mn}$ and $B_{m\nb}$. We stress
again that eq.~\refs{m_mdub} has been derived taking the full $\DH =11$
spacetime into account. 

We start to explore the invariances of eq.~\refs{m_mdub} by considering
the vertex operator transformations
\be
 \wtd{\cF}\rightarrow P\wtd{\cF}\; ,\quad \cF\rightarrow {P^{-1}}^T\cF
  \label{m_Ftrans}
\ee
and the moduli transformation
\be
 M\rightarrow PMP^T\; ,
 \label{m_Mtrans}
\ee
where $P\in \bG\subset GL(d(d+1)/2)$. Clearly, the
transformation~\refs{m_Mtrans} should preserve the structure of $M$. This
restricts the allowed matrices $P$ and determines the group $\bG$. A useful
general piece of information can be extracted from the formula
\be
 \mbox{det}(M) = \F^{\frac{3}{2}\frac{(D-7)(D-8)}{D-2}}
 \label{m_det}
\ee
which follows from the definition of $M$, eq.~\refs{m_M}. It shows that
$\mbox{det}(M) = 1$ for $D=7,8$ so that $P\in \bG\subset SL(d(d+1)/2)$
in these dimensions. Let us now analyze the structure of $M$ case by case.

\vspace{0.4cm}

$\underline{D=10}$~: In this case eq.~\refs{m_mdub} is a 1--dimensional
equation with
\be
 M=\left( \F^{9/8}\right)\; .
\ee
The duality group $\bG = GL(1)$ coincides with the internal global
coordinate transformations and extends therefore trivially to a symmetry
of the full membrane equations (including all background fields). On the
level of the effective theory it corresponds to the scaling symmetry
of type IIA supergravity.

\vspace{0.4cm}

$\underline{D=9}$~: Eq.~\refs{m_mdub} is now a 3--component vector
equation with the first two components labeled by $m=9,10$ and the
third component by an antisymmetric index pair $\mb$. The matrix $M$
contains $3$ (metric) moduli. We can use the internal $\e$ tensor
$\e^\mb$ for a relabeling of basis vectors by contracting it with
index pairs $\mb$. Therefore we define modified vertex operators by
\be
 \wtd{\cF} = \left(\ba{c} \wtd{\cF}^{(1)}_m \\ 
                          \frac{1}{\F}\e_\mb\wtd{\cF}^{(2)\mb}\ea\right)\; ,
                          \quad
 \cF = \F^{-1/7}\left(\ba{c} \cF^{(1)m} \\ \e^\mb\cF^{(2)}_\mb\ea\right)\; .
\ee
For the correspondingly transformed matrix $M$ one finds
\be
 M = \left(\ba{cc} \F^{1/7}G_{mn} & 0 \\ 0 & \F^{-6/7} \ea\right)\; .
\ee
Since $\mbox{det}(M)\neq 1$ in accordance with the general formula~\refs{m_det}
the duality group is $\bG = GL(2)$. Again this coincides with the
internal global coordinate transformations and therefore extends trivially
to the full theory. We observe that the vertex operators $\wtd{\cF}$, $\cF$
transform in the reducible representation ${\bf 2}+{\bf 1}$ of $GL(2)$ so
that the conjugate momenta in the first two components and the Bianchi
identity in the third component do not mix. The matrix $M$
parameterizes the moduli coset $GL(2)/SO(2)$. Clearly, the same $GL(2)$
transformations are found as the U--duality group of $D=9$ supergravity.

\vspace{0.4cm}

$\underline{D=8}$~: Now eq.~\refs{m_mdub} represents a 6--component
vector equation where the first three entries are labeled by $m=8,9,10$ and
the last three components by an antisymmetric pair $\mb$. The moduli
space is 7--dimensional with 6 metric moduli and 1 modulus from the 3--form.
We can apply a similar method as in the $D=9$ case and use the internal
$\e$ symbol $\e^{m\mb}$ to convert all index pairs $\mb$ into single indices.
Then the vertex operators get modified to
\be
 \wtd{\cF} = \left(\ba{c} \wtd{\cF}^{(1)}_m \\ 
                          \frac{1}{\F}\e_{m\mb}\wtd{\cF}^{(2)\mb}\ea\right)\; ,
                          \quad
 \cF = \F^{-1/6}\left(\ba{c} \cF^{(1)m} \\ \e^{m\mb}\cF^{(2)}_\mb\ea\right)\; .
 \label{m_F8}
\ee 
Note that now the upper and lower component of $\wtd{\cF}$, $\cF$ have the
same index structure. This reflects the earlier mentioned fact that the
number of conjugate momenta equals the number of winding modes (of both spatial
worldvolume directions) in $D=8$ (and only in $D=8$). In the basis~\refs{m_F8}
the matrix $M$ takes the form
\be
 M = M_2\otimes M_3
\ee
with
\be
 M_2 = \F^{1/2}\left(\ba{cc} 1+\frac{B^2}{\F} & \frac{B}{\F} \\
                              \frac{B}{\F} & \frac{1}{\F}\ea\right)\; ,
       \quad
 M_3 = \F^{-1/3}\left( G_{mn}\right)\; . \label{m_M23}
\ee
Here $B$ is the single modulus from the 3--form defined by
\be
 B_{m\nb} = \frac{B}{\F}\e_{m\nb}\; .
\ee
From the tensor structure of $M$, and the fact that
$\mbox{det}(M_2)=\mbox{det}(M_3)=1$, we learn that the duality group in this
case is given by $\bG = SL(2)\times SL(3)$. This group is indeed the known
U--duality group of $D=8$ supergravity. The matrix $M$ represents a
parameterization of the moduli coset $SL(2)/SO(2)\times SL(3)/SO(3)$.

We would like to be more specific about the action of a group element
$P\in\bG$. We therefore split $P=P_2\otimes P_3$ with $P_2$, $P_3$
in the defining representations of $SL(2)$, $SL(3)$. The two parts of $M$
then transform as
\be
 M_{2,3}\rightarrow P_{2,3}M_{2,3}P_{2,3}^T\; .
 \label{m_trans8}
\ee
The action on the vertex operators~\refs{m_F8} is described as follows.
The $SL(3)$ transformation $P_3$ acts on the internal index $m$ in
eq.~\refs{m_F8}, simultaneously for the upper and lower components
(the conjugate momenta and Bianchi identities). This part of the group
therefore consists of global internal coordinate transformations and
extends trivially to the full equations of motion. The situation is
quite different for the $SL(2)$ part. It acts on the upper and lower
component of $\wtd{\cF}$, $\cF$ (for each $m$ in the same way) and therefore
exchanges momentum and winding modes of the membrane. In this sense, it appears
to be the direct analog of a string T--duality transformation. On the other
hand, let us consider the specific $SL(2)$ transformation
\be
 S = \left(\ba{cc} 0&1\\-1&0\ea\right)\in SL(2)
\ee
and let the 3--form modulus $B=0$. Then from eq.~\refs{m_trans8} this
transformation acts on the dilaton as
\be
 \F\stackrel{S}{\rightarrow}\frac{1}{\F}\; ,
\ee
that is, as an S--duality transformation. This confirms the general expectation
that S duality should arise as a momentum/winding--mode exchange on the
membrane worldvolume since the dilaton is just a geometrical modulus within
the framework of M theory. We remark that the same $SL(2)$ symmetry, acting
on the moduli of the theory, can be found within the framework of the matrix
model quantization of M--theory~\cite{wati}. It is by no means obvious that
the $SL(2)$ symmetry can be extended to the full membrane equations of motion
including all background fields. This question will be studied in detail
in the next section.

\vspace{0.4cm}

$\underline{D=7}$~: Eq.~\refs{m_mdub} represents a 10--component vector
equation with the first four entries indexed by $m=7,8,9,10$ and the
others by an antisymmetric pair $\mb$. There are 14 moduli, 10 from the
metric and 4 from the antisymmetric tensor field. As in the previous
examples, the internal $\e$ tensor can be used to relabel the basis vectors
such that the duality symmetry becomes manifest. Since the moduli
equation~\refs{m_mdub} is similar in form to the one found by Duff and
Lu~\cite{dl}, our results for the transformation of  $\wtd{\cF}$, $\cF$ and
$M$ coincide with the ones given there. We will therefore not give the
explicit formulae here, but refer to ref.~\cite{dl} instead. Let us
just mention that the duality group is $G=SL(5)$ which coincides with
the U--duality group in $D=7$. Under this group, the vertex operators
$\wtd{\cF}$, $\cF$ transform as the second rank antisymmetric tensor
representation ${\bf 10}$ and $M$ parameterizes the coset $SL(5)/SO(5)$. 

\vspace{0.4cm}

$\underline{D=6}$~: In this dimension, for the first time, we encounter a
paradox. Eq.~\refs{m_mdub} is a 15--component equation with the first
five components indexed by $m=6,7,8,9,10$ and the rest by an antisymmetric
pair $\mb$. The known U--duality group in this dimension is $O(5,5)$ and
there is obviously no 15--dimensional representation of this group under
which $\wtd{\cF}$, $\cF$ could transform. A resolution of this paradox
comes from the observation that, within the framework of the low energy
effective action, antisymmetric $\d$ forms should be Poincar\'e dualized
to $D-\d -2$ forms if $\d >(D-2)/2$ (if $\d =(D-2)/2$ the $\d$ form should
be paired in the Gaillard--Zumino way~\cite{gz}) in order to discover the
full U--duality group.
Therefore, in $D=6$ we should dualize the 3 form $\bh_{\m\n\r}$ to a
vector field. Instead of 15 vector fields in ${\bf A}_\n$, eq.~\refs{m_vec},
we are now dealing with 16 which then transform under the spinor
representation of $O(5,5)$. Since ${\bf A}_\n$ has the same internal index
structure as $\cF$ and enters the full internal equations of
motion~\refs{m_dubl} in a similar way, it seems natural that eq.~\refs{m_dubl}
should be augmented by one component and should transform as a spinor under
$O(5,5)$. Then, also $\wtd{\cF}$, $\cF$ would form spinor representations
of $O(5,5)$. Though such a Poincar\'e dualization is straightforwardly
performed in the low energy effective action, it is unclear (to us) how
this can be done for the worldvolume theory. Despite the fact that we
have no ``missing multiplets'' (as this phenomenon was called in
ref.~\cite{dl}), since we have taken the full background including
$\bh_{\m\n\r}$ into account, we remain unable to find a manifestly
$SL(5)$ invariant form of the moduli part in the $D=6$ case.

It is clear that the problem of ``Poincar\'e dualizing on the worldvolume''
also arises in other dimensions and affects more fields the lower the
dimension is. However, even in $D=7$ we would have encountered this
problem if we had taken the 2 forms $\bh_{\m\n m}$ into account. These
2 forms fit into a multiplet of the $D=7$ U--duality group $SL(5)$ only
if they are augmented by the dual of the 3 form $\bh_{\m\n\r}$. In $D=8$
the dualizing problem affects the 3 form $\bh_{\m\n\r}$ only. In the
low energy effective action it has to be paired in the Gaillard--Zumino
way to form an $SL(2)$ doublet. All other background fields, however, fit
into multiplets of the $D=8$ U--duality group $SL(2)\times SL(3)$ without
dualization. It is for this reason that we will concentrate on the $D=8$
example in the next section.
 
\section{The Example $D=8$ with General Background}
In the previous section, we have discussed how U--duality symmetries can
be read off from the moduli part of the worldvolume theory. Here, we would
like to generalize this discussion to include the full content of background
fields. As we have seen, this generalization is trivial in $D=9,10$ since
the U--duality groups in these dimensions coincide with the global internal
coordinate transformations. On the other hand, if we decrease the dimension
some background fields have to be Poincar\'e dualized to discover the full
U--duality group and, unfortunately, we are generally unable to do this in the
worldvolume theory. The ``cleanest'' case, from this point of view, is
the $D=8$ one, as explained in the end of the last section. Given our
ignorance on how to perform the dualization explicitly on the worldvolume,
we therefore concentrate on the $D=8$ example.

First, we would like to analyze the internal equations of motion,
eq.~\refs{m_dubl1}, in this case. They read
\be
 \wtd{\cF} = M\left( \cF+{\bf A}_\n{\cal A}^\n+{\bf V}_{r\n}{\cal V}^{r\n}+
             \frac{1}{2}{\bf B}_{\n\s}{\cal B}^{\n\s}
             \right)\; .
 \label{m_dubl2}
\ee
with the vertex operators $\wtd{\cF}$, $\cF$ and ${\cal A}^\n$,
${\cal V}^{r\n}$, ${\cal B}^{\n\s}$ defined in eq.~\refs{m_F} and
\refs{m_AVB} and the background $M$, ${\bf A}_\n$, ${\bf V}_{r\n}$
and ${\bf B}_{\n\s}$ defined in eq.~\refs{m_M}, \refs{m_vec} and
\refs{m_twof}. In the previous section, we have already analyzed the moduli
part, $\wtd{\cF} = M\cF$, of this equation. It turned out that the internal
$\e$ symbol $\e^{m\mb}$ should be used to convert antisymmetric index pairs
$\mb$ into single indices $m$. The vertex operators relabeled in such a way
then read
\be
 \wtd{\cF}_m = \left(\ba{c} \wtd{\cF}^{(1)}_m \\ 
                          \frac{1}{\F}\e_{m\mb}\wtd{\cF}^{(2)\mb}\ea\right)\; ,
                          \quad
 \cF^m = \F^{-1/6}\left(\ba{c} \cF^{(1)m} \\
          \e^{m\mb}\cF^{(2)}_\mb\ea\right)\; ,
 \label{m_F81}
\ee 
where we have made the internal index $m$ explicit in this notation.
Correspondingly, the moduli matrix $M$ in this basis takes the form
\be
 M=M_2\otimes M_3
\ee
with $M_2$, $M_3$ given in eq.~\refs{m_M23}. Since
$\mbox{det}(M_2)=\mbox{det}(M_3)=1$ this structure of $M$ determines the
group of duality rotations to be ${\bf G}=SL(2)\times SL(3)$. A group
element $P=P_2\otimes P_3$ with $P_2$, $P_3$ in the defining representation
of $SL(2)$, $SL(3)$ acts on $\wtd{\cF}$, $\cF$ as
\be
 \wtd{\cF}_m \rightarrow \left( P_3\right)_m^{m'}P_2\wtd{\cF}_{m'}\; ,
 \quad
  \cF^m \rightarrow \left( {P_3^{-1}}^T\right)^m_{m'}{P_2^{-1}}^T\cF^{m'}\; ,
 \label{m_Fex}
\ee
and on $M$ as
\be
 M_{2,3}\rightarrow P_{2,3}M_{2,3}P_{2,3}^T\; .
 \label{m_Mtrans1}
\ee
Since $SL(3)$ is part of the global internal coordinate transformations,
$P_3$ generally transforms the internal indices. In particular this is true
for $\wtd{\cF}$, $\cF$. An $SL(2)$ transformation
$P_2$, on the other hand, acts on the upper and lower component of
$\wtd{\cF}$, $\cF$ (similarly for each $m$) and therefore exchanges
momentum and winding modes. As we have seen, $SL(2)$ is an S--duality symmetry
which, in particular, contains the dilaton transformation
$\F\rightarrow 1/\F$.

\vspace{0.4cm}

Let us now extend this picture to the other background fields. We begin
with the vector fields in ${\bf A}_\n$, ${\bf V}_{r\n}$. In the
basis~\refs{m_F81} these vectors read
\be
 {\bf A}_\n^m = \left(\ba{c} A_\n^{(1)m}\\ \e^{m\mb}A_{\n\mb}^{(2)}\ea\right)
 \; ,\quad
 {\bf V}_{r\n}^m = \F^{-1/6}\left(\ba{c} V_{r\n}^{(1)m}\\
                 \e^{m\mb}V_{r\n\mb}^{(2)}\ea\right)\; ,
 \label{m_vec1}
\ee
where we have made the $SL(3)$ index $m$ explicit in the notation. Let us
first discuss the transformation of the corresponding vertex operators
${\cal A}^\n$ and ${\cal V}^{r\n}$. As in the string case, from their
index structure, we expect them to be singlets under $SL(2)$. On the other
hand, since ${\cal V}^{r\n}$ carries an internal index, it transforms under
$SL(3)$. Therefore we start with
\be
 {\cal A}^\n\rightarrow {\cal A}^\n\; ,\quad {\cal V}^{r\n}\rightarrow
   \left( {P_3^{-1}}^T\right)^r_{r'}{\cal V}^{r' \n}
 \label{m_AVtrans}
\ee
as the $SL(2)\times SL(3)$ transformation law for the vertex operators. Given
the transformation of $\cF$ in eq.~\refs{m_Fex}, and the structure of the
internal equations of motion~\refs{m_dubl2}, this forces us to require the
following transformations for the vectors ${\bf A}_\n^m$ and ${\bf V}_{r\n}^m$
\be
 {\bf A}_\n^m\rightarrow \left( {P_3^{-1}}^T\right)^m_{m'}P_2^{-1T}
       {\bf A}_\n^{m'}\; ,\quad
 {\bf V}_{r\n}^m\rightarrow \left( {P_3^{-1}}^T\right)^m_{m'}
    \left( P_3\right)_r^{r'}{P_2^{-1}}^TV_{r' \n}^{m'}\; .
 \label{m_vtrans}
\ee
Observe that both vectors are $SL(2)$ doublets so that
graviphotons and vector fields from the antisymmetric tensor are rotated into
each other. As for the string, ${\bf A}_\n^m$ and ${\bf V}_{r\n}^m$
contain the same degrees of freedom, namely the six vector fields of the
theory, and one therefore has to check the consistency of the two
transformations~\refs{m_vtrans}. From eq.~\refs{m_vec1} and the
definitions~\refs{m_fields} we find
\be
 {\bf V}^m_{r\n} = \e^{muv}M_{3ru}M_{3vn}\e_2M_2{\bf A}_\n^n
\ee
with
\be
 \e_2 = \left(\ba{cc} 0 & 1 \\ -1 & 0 \ea\right)\; ,
\ee
which is the analog of eq.~\refs{st_vcons} for the string. Using the
transformations~\refs{m_Mtrans1} for $M$ and ~\refs{m_vtrans} for
${\bf A}_\n^m$, together with $P_2^T\e_2 P_2 = \e_2$, shows the consistency  
of eq.~\refs{m_vtrans}. Therefore, the vector field terms in the internal
equations of motion~\refs{m_dubl2} (the second and third term on the right
hand side) are compatible with the $SL(2)\times SL(3)$ symmetry.
The transformations~\refs{m_vtrans} for the vector fields exactly coincide
with the ones obtained from the low energy effective action, as we will
see in the next section.

Finally, to establish an invariance of the internal equations of motion, we
need to consider the vector ${\bf B}_{\n\s}$ which, in the
basis~\refs{m_F81}, reads
\be
 {\bf B}_{\n\s}^m = \left(\ba{c} B_{\n\s}^{(1)m} \\ \e^{m\mb}B_{\n\s\mb}^{(2)}
                    \ea\right)\; .
 \label{m_Bvec}
\ee
The ``natural'' assumption for the corresponding vertex operator
${\cal B}^{i\n\s}=\e^{ijk}\pa_jX^\n\pa_kX^\s$ is that it transforms as a
singlet under $SL(2)\times SL(3)$. From eq.~\refs{m_dubl2}, this requires
${\bf B}_{\n\s}^m$ to be an $SL(2)$ doublet. On the other hand, let us recall
from eq.~\refs{m_fields} the definition of $B_{\n\s}^{(1)m}$ and
$B_{\n\s\mb}^{(2)}$
\bea
 B_{\n\s}^{(1)m} &=& \gh^{mn}\bh_{n\n\s} \\
 B^{(2)}_{\n\s\mb} &=& -\gh_{\mb\n\s}-\bh_{n\mb}B_{\n\s}^{(1)n} \nn \\
                   &=& -A_{\n m_1}^{(1)}A_{\s m_2}^{(1)}+
                        A_{\n m_2}^{(1)}A_{\s m_1}^{(1)}-
                        \bh_{n\mb}B_{\n\s}^{(1)n}\; .
\eea
We know already that $A^{(1)}_{\n m}$ is the upper component of an $SL(2)$
doublet. Therefore, as the second expression above contains bilinears in
$A^{(1)}_{\n m}$, the lower entry $B^{(2)}_{\n\s\mb}$ of the
vector~\refs{m_Bvec} cannot transform as part of a doublet in contradiction
to our previous assumption. Moreover, the new degrees of freedom in
${\bf B}_{\n\s}^m$ are the 2 forms $\bh_{n\n\s}$, and it is well known from
$D=8$ supergravity that they are $SL(2)$ singlets. The only possible
conclusion, at this point, therefore is, that the last term in
eq.~\refs{m_dubl2} breaks the $SL(2)$ invariance of the internal equations
of motion. One might argue that eq.~\refs{m_dubl2}, though the correct
equation of motion, is not in an appropriate form to manifestly show
the $SL(2)$ invariance. Unfortunately, all our attempts to remove the
obstruction by modifying the form of eq.~\refs{m_dubl2} failed. We will
comment on these attempts and on possible reasons for the symmetry
breaking in the final section 7.

\vspace{0.4cm}

To complete the picture, we would now like to analyze the other equations
of motion as well. Eq.~\refs{m_single1} for the mixed Bianchi identity
can be written as
\be
 {\cal V}^{m\n} = -\left( M_3^{-1}\right)^{mn}g^{\n\r}\left[
                  \frac{1}{\F}\e_{nrs}{\cF^r}^T\e_2{\bf A}_\r^s +
                  \left( B_{\r\s n}+\frac{1}{2\F}\e_{nrs}{{\bf A}_\r^r}^T
                  \e_2{\bf A}_\s^s\right){\cal A}^\s - {\cal J}_{n\r}\right]
                  -A_\r^{(1)m}{\cal B}^{\r\n}\; .
 \label{m_single2}
\ee
Let us assign the transformation property
\be
 {\cal J}_{n\s}\rightarrow\left( P_3\right)^{n'}_n {\cal J}_{n' \s}
 \label{m_Jtrans}
\ee
to the vertex operator ${\cal J}_{n\s}$ and the following transformations
to the metric~\refs{m_g} and the 2 forms~\refs{m_B}
\bea
 g_{\m\n}&\rightarrow& g_{\m\n} \label{m_metric} \\
 B_{\n\s n} &\rightarrow& \left( P_3\right)_n^{n'}B_{\n\s n'}\; .
 \label{m_2trans}
\eea
Then, with the transformations~\refs{m_Mtrans}, \refs{m_AVtrans}
and \refs{m_vtrans}, eq.~\refs{m_single2} is invariant under
$SL(2)\times SL(3)$ up to the last term. It is interesting to observe that
the symmetry breaking term is again proportional to the vertex operator
${\cal B}^{i\n\s}=\e^{ijk}\pa_jX^\n\pa_kX^\s$ which is bilinear in the
external coordinates $X^\m$. Note that, though we could not read off the
correct 2 form transformation law form the internal equations of
motion~\refs{m_dubl2}, we could do so from eq.~\refs{m_single2}. As we will
verify in the next section, the transformation laws~\refs{m_metric} and
\refs{m_2trans} are indeed correct.
                  
\vspace{0.4cm}

Finally, we analyze the external conjugate momentum~\refs{m_sing} which now
reads
\bea
 \frac{\pa{\cal L}_x}{\pa (\pa_iX^\m )} &=&
    {\bf A}^T_\m{\wtd{\cF}}^i+g_{\m\n}{\cal A}^{i\n}+
    \left( B_{\n\m r}+\frac{1}{2\F}\e_{rst}{{\bf A}_\n^s}^T\e_2
    {\bf A}_\m^t\right){\cal V}^{ir\n} \nn \\
    && +\frac{1}{2}(\bh_{\m\n\r}-{\gh^r }_\m\bh_{\n\r r}){\cal B}^{i\n\r}\; .
    \label{m_sing1}
\eea
Again, all terms except the one proportional to ${\cal B}^{i\n\s}$ are
$SL(2)\times SL(3)$ invariant. In eq.~\refs{m_sing1} this term is
associated with the 3 form $\bh_{\m\n\r}$. Therefore, this symmetry
breaking is no surprise, since, as noted earlier, the 3 form has to be paired
in the Gaillard--Zumino way to form an $SL(2)$ doublet. Because we did not do
this in the worldvolume theory, the obstruction in eq.~\refs{m_sing1} is an
expected one. It is conceivable that the origin of the other two symmetry
breaking terms in eq.~\refs{m_dubl2} and \refs{m_single2} is related to this.

\vspace{0.4cm}

To summarize, from the membrane equations of motion~\refs{m_dubl2},
\refs{m_single2}, \refs{m_sing1} in $D=8$ we have read off the
$SL(2)\times SL(3)$ transformations which act on the vertex operators
as in eq.~\refs{m_Fex}, \refs{m_AVtrans}, \refs{m_Jtrans} and on the
background fields as in eq.~\refs{m_Mtrans1}, \refs{m_vtrans},
\refs{m_metric}, \refs{m_2trans}. Unfortunately, these transformations
do not constitute a symmetry of the equations of motion, but leave all
terms except the ones proportional to the vertex operator
${\cal B}^{i\n\s}=\e^{ijk}\pa_jX^\n\pa_kX^\s$ invariant. Despite this
fact, the background field transformations agree exactly with the results from
$D=8$ supergravity, as we will show in the next section.

\section{Comparison with $D=8$ Supergravity}

In this section, we would like to verify the background transformation
laws as determined from the membrane worldvolume theory. This will be
done by comparison with $\DH =11$ supergravity dimensionally reduced to
$D=8$~\cite{sal_sez}. Though we are mainly interested in the specific
dimension $D=8$, the first part of the dimensional reduction procedure
will be kept general.

The bosonic part of the $\DH =11$ supergravity Lagrangian
reads~\footnote{We are using the conventions of ref.~\cite{cj} except for
an additional rescaling of the 3 form by a factor $1/2$.}
\be
 {\cal L}_{11} = \sqrt{-\gh}\left[\frac{1}{4}\hat{R}-\frac{1}{8\cdot 4!}
                 \hat{F}_{MNPQ}\hat{F}^{MNPQ}\right]
                 +\frac{1}{82944}\e^{M_1...M_{11}}\hat{F}_{M_1...M_4}
                  \hat{F}_{M_5...M_8}\bh_{M_9M_{10}M_{11}}
 \label{L11}
\ee
with
\be
 \hat{F} = 4\pa_{[M}\bh_{NPQ]}\; .
\ee
Our index convention for the dimensional reduction is the same as in the
previous sections. We use indices $M,N,P,...=0,...,10$ for the full space,
indices $\m ,\n ,\r ,...=0,...,D-1$ for the external space and indices
$m,n,r,...=D,...,10$ for the internal space. For each of these index types
we will need corresponding flat tangent space indices which we denote by
$A,B,C,...=0,...,10$ for the full space, $\a ,\b ,\g ,...=0,...,D-1$
for the external space and $a,b,c,...=D,...,10$ for the internal space.
For our purpose, it will be sufficient to consider the non--topological
part of the Lagrangian~\refs{L11}; that is,
\be
 {\cal L}_0 = \sqrt{-\gh}\left[\frac{1}{4}\hat{R}-\frac{1}{8\cdot 4!}
                 \hat{F}_{MNPQ}\hat{F}^{MNPQ}\right] \; .
 \label{L0}
\ee
Following standard methods for dimensional reduction~\cite{dr} we use
the Ansatz
\be
 \hat{e}^A_M = \left(\ba{cc} \bar{e}_\m^\a & A^{(1)n}_\m E^a_n \\
                             0 & E^a_m\ea\right)
\ee
for the vielbein $\hat{e}^A_M$ with
$\gh_{MN} = \eta_{AB}\hat{e}^A_M\hat{e}^B_N$. The internal and external
metrics are defined by
\bea
 G_{mn} &=& \d_{ab}E^a_mE^b_n \\
 \bar{g}_{\m\n} &=& \eta_{\a\b}\bar{e}^\a_\m\bar{e}^\b_\n\; ,
\eea
respectively. For the total metric $\gh_{MN}$ we then get
\be
 \gh_{MN} = \left(\ba{cc} \bar{g}_{\m\n}+A^{(1)}_{\m r}A^{(1)r}_\n &
                          A^{(1)}_{\m n} \\
                          A^{(1)}_{\n m} & G_{mn}\ea\right)\; .
\ee
The most convenient way to perform the dimensional reduction of the 3 form
kinetic term in the Lagrangian~\refs{L0} is to first express it in terms
of the flat field strength
$\hat{F}_{ABCD} = \hat{e}^M_A\hat{e}^N_B\hat{e}^P_C\hat{e}^Q_D\hat{F}_{MNPQ}$,
then perform the reduction and finally convert back to curved external
indices using $\bar{e}^\a_\m$. In such a way, and by inserting the
above expression for the metric, we arrive at the following Lagrangian
\bea
 {\cal L}_0 &=& \sqrt{-\bar{g}}\;\sqrt{\F}\left[\frac{1}{4}\bar{R}+\frac{1}{16}
                \pa_\m G_{mn}\pa^\m G^{mn}+\frac{1}{16}\F^{-2}\pa_\m\F
                \pa^\m\F \right. \nn \\
            &&\quad\quad\quad\quad\quad
                -\frac{1}{16}G_{mn}F^{(1)m}_{\m\n}F^{(1)\m\n n}
              -\frac{1}{8\cdot 4!}F_{\m\n\r\s}F^{\m\n\r\s}
              -\frac{1}{8\cdot 3!}F_{\m\n\r s}F^{\m\n\r s}\nn \\
            &&\quad\quad\quad\quad\quad
               \left. -\frac{1}{32}F_{\m\n rs}F^{\m\n rs}
               -\frac{1}{8\cdot 3!}F_{\m nrs}F^{\m nrs}\right]
 \label{red1}
\eea
with the dilaton
\be
 \F = \mbox{det}(G_{mn})
\ee
and
\bea
 F^{(1)m}_{\m\n} &=& \pa_\m A^{(1)m}_\n -\pa_\n A^{(1)m}_\m \nn \\
 F_{\m\n\r\s} &=& \bar{e}_\m^\a \bar{e}_\n^\b \bar{e}_\r^\g \bar{e}_\s^\d
                  \hat{e}^M_\a \hat{e}^N_\b \hat{e}^P_\g \hat{e}^Q_\d
                  \hat{F}_{MNPQ} \nn \\
 F_{\m\n\r s} &=& \bar{e}_\m^\a \bar{e}_\n^\b \bar{e}_\r^\g
                  \hat{e}^M_\a \hat{e}^N_\b \hat{e}^P_\g\hat{F}_{MNPs} 
                  \label{fst0} \\
 F_{\m\n rs} &=& \bar{e}_\m^\a \bar{e}_\n^\b\hat{e}^M_\a \hat{e}^N_\b
                  \hat{F}_{MNrs} \nn \\
 F_{\m nrs} &=& \bar{e}_\m^\a\hat{e}^M_\a\hat{F}_{Mnrs} \nn\; .
\eea
To get rid of the factor $\sqrt{\F}$ on the right hand side on eq.~\refs{red1},
we perform a Weyl rescaling of the external metric $\bar{g}_{\m\n}$ to
\be
 g_{\m\n} = \F^{\frac{1}{D-2}}\bar{g}_{\m\n} =
             \F^{\frac{1}{D-2}}\left(\gh_{\m\n}-
            A^{(1)r}_\m A^{(1)}_{\n r}\right)\; .
 \label{metric}
\ee
Furthermore, we split ${\cal L}_0$ into a gravitational, a moduli and
a form part as
\be
 {\cal L}_0 = {\cal L}_{\rm gr}+{\cal L}_{\rm moduli}+
              {\cal L}_{\rm 1\; forms}+{\cal L}_{\rm 2\; forms}+
              {\cal L}_{\rm 3\; form}\; .
\ee
For these various parts we find
\bea
 {\cal L}_{\rm gr} &=& \sqrt{-g}\frac{1}{4}R \nn \\
 {\cal L}_{\rm moduli} &=& \sqrt{-g}\left[ -\frac{1}{16(D-2)}\F^{-2}
                              \pa_\m\F\pa^\m\F+\frac{1}{16}\pa_\m G_{mn}
                              \pa^\m G^{mn}\right.\nn \\
                       &&\left.\quad\quad\quad -\frac{1}{8\cdot 3!}G^{mm'}
                              G^{nn'}G^{rr'}\pa_\m B_{mnr}\pa^\m B_{m' n' r'}
                              \right] \nn \\
 {\cal L}_{\rm 1\; forms} &=& \sqrt{-g}\left[ -\frac{1}{16}\F^{\frac{1}{D-2}}
                               G_{mn}F^{(1)m}_{\m\n}F^{(1)\m\n n}
                               -\frac{1}{32}\F^{\frac{1}{D-2}}F_{\m\n rs}
                               F^{\m\n rs}\right] \label{Lparts}\\
 {\cal L}_{\rm 2\; forms} &=& \sqrt{-g}\left[ -\frac{1}{8\cdot 3!}
                               \F^{\frac{2}{D-2}}F_{\m\n\r s}F^{\m\n\r s}
                               \right] \nn \\
 {\cal L}_{\rm 3\; form} &=& \sqrt{-g}\left[ -\frac{1}{8\cdot 4!}
                               \F^{\frac{3}{D-2}}F_{\m\n\r\s}F^{\m\n\r\s}
                               \right] \nn\; ,
\eea
where, from eq.~\refs{fst0}, the field strengths are given by
\bea
 F^{(1)m}_{\m\n} &=& \pa_\m A^{(1)m}_\n -\pa_\n A^{(1)m}_\m \nn \\
 F_{\m\n rs} &=& \hat{F}_{\m\n rs}-A^{(1)m}_\m\hat{F}_{m\n rs}
                 -A^{(1)n}_\n\hat{F}_{\m nrs} \nn \\
 F_{\m\n\r s} &=& \hat{F}_{\m\n\r s}-\left( A_\m^{(1)m}\hat{F}_{m\n\r s}+
                  \mbox{2 perm}\right) +
                  \left( A^{(1)m}_\m A^{(1)n}_\n\hat{F}_{mn\r s}+
	          \mbox{2 perm}\right) \label{fst1}\\
 F_{\m\n\r\s} &=& \hat{F}_{\m\n\r\s}-\left(A^{(1)m}_\m\hat{F}_{m\n\r\s}+
                  \mbox{3 perm}\right) +
                  \left( A^{(1)m}_\m A^{(1)n}_\n\hat{F}_{mn\r\s}+
                  \mbox{5 perm}\right) \nn \\
              && - \left( A^{(1)m}_\m A^{(1)n}_\n A^{(1)r}_\r\hat{F}_{mnr\s}+
                  \mbox{2 perm}\right)\nn \; .
\eea
So far, we have kept the dimension $D$ general. In what follows, we will
concentrate on the case $D=8$ to show that the Lagrangian ${\cal L}_0$
(except ${\cal L}_{\rm 3\; form}$, see the discussion below) is invariant
under the $SL(2)\times SL(3)$ background transformations which we have read 
off from the membrane worldvolume theory in the previous section. It turns
out that the various parts of ${\cal L}_0$, as listed in eq.~\refs{Lparts},
(again except ${\cal L}_{\rm 3\; form}$) are independently invariant
under these transformations. Therefore, we discuss each of these parts
separately. We start with

\vspace{0.4cm}

$\underline{{\cal L}_{\rm gr}}$~: This part of the Lagrangian only
depends on the Weyl rescaled metric $g_{\m\n}$, eq.~\refs{metric}, which,
according to eqs.~\refs{m_g} and \refs{m_metric} is a singlet under
$SL(2)\times SL(3)$.

\vspace{0.4cm}

$\underline{{\cal L}_{\rm moduli}}$~: To show the $SL(2)\times SL(3)$
invariance, we would like to express ${\cal L}_{\rm moduli}$ in terms
of the covariantly transforming quantities $M_2$, $M_3$ defined in
eq.~\refs{m_M23}. A straightforward computation using these definitions leads
to
\be
 {\cal L}_{\rm moduli} = \sqrt{-g}\frac{1}{16}\left[\mbox{tr}\left(
                            \pa_\m M_2\pa^\m M_2^{-1}\right) +
                            \mbox{tr}\left( \pa_\m M_3\pa^\m M_3^{-1}\right)
                            \right]\; ,
\ee
which is invariant under the $SL(2)\times SL(3)$
transformations~\refs{m_Mtrans1} and \refs{m_metric}.

\vspace{0.4cm}

$\underline{{\cal L}_{\rm 1\; forms}}$~: This part of the Lagrangian contains
the vector fields, the moduli and the external metric. It should be expressible
in terms of the vector ${\bf A}_\n$ in eq.~\refs{m_vec1} and the matrix
$M=M_2\otimes M_3$. Indeed, from eq.~\refs{m_vec1}, \refs{m_M23} and
\refs{m_fields} we find
\be
 {\cal L}_{\rm 1\; forms} = -\frac{1}{16}\sqrt{-g}\;{\bf F}_{\m\n}^T
                             M{\bf F}^{\m\n}\; ,
\ee
where
\be
 {\bf F}_{\m\n} = \pa_\m{\bf A}_\n -\pa_\n{\bf A}_\m
\ee
is the $SL(2)\times SL(3)$--covariant field strength. This is, in fact,
manifestly invariant under the transformations~\refs{m_Mtrans1},
\refs{m_vtrans} and \refs{m_metric}.

\vspace{0.4cm}

$\underline{{\cal L}_{\rm 2\; forms}}$~: This part contains the 2 forms,
the vector fields, the moduli and the external metric. We expect the
relevant covariant quantities to be $B_{\m\n r}$ in eq.~\refs{m_B},
${\bf A}_\n$ in eq.~\refs{m_vec1} and the matrix $M$. After some
computation we find that the field strength $F_{\m\n\r s}$ in
eq.~\refs{fst1} can be written as
\be
 F_{\m\n\r s} = H_{\m\n\r s} -\frac{1}{2}\left(\frac{1}{\F}\e_{smn}
                {{\bf A}_\m^m}^T\e_2{\bf F}_{\n\r}^n+\mbox{2 perm}\right)\; ,
\ee
where
\be
 H_{\m\n\r s} = 3\pa_{[\m}B_{\n\r ]s}
\ee
is the 2 form field strength. The Lagrangian ${\cal L}_{\rm 2\; forms}$
then takes the form
\be
 {\cal L}_{\rm 2\; forms}=-\frac{1}{8\cdot 3!}\left( M_3^{-1}\right)^{rs}
                            F_{\m\n\r r}{F^{\m\n\r}}_s\; .
\ee
Under the transformations~\refs{m_2trans}, \refs{m_vtrans} and \refs{m_Mtrans1}
the field strength $F_{\m\n\r s}$ is $SL(2)$--invariant and $SL(3)$--covariant.
This shows the invariance of ${\cal L}_{\rm 2\; forms}$.

\vspace{0.4cm}

$\underline{{\cal L}_{\rm 3\; form}}$~: It is well known that this part
of the Lagrangian is not invariant under $SL(2)$ by itself. Instead, the
$SL(2)$ symmetry shows up as a Gaillard--Zumino duality rotation between
the equations of motion and the Bianchi identity of the 3 form. This
means that the 4--form field strength $D_{\m\n\r\s}=4\pa_{[\m}C_{\n\r\s ]}$
(which consists of $\bh_{\m\n\r}$ and lower degree forms) has to be
paired with $\frac{\pa {\cal L}_{11}}{\pa D_{\m\n\r\s}}$ to form an $SL(2)$
doublet. As we have already pointed out, we do not know how to perform such
a Gaillard--Zumino duality transformation of background fields in the
worldvolume theory. We were therefore unable to derive the 3 form transformation
law from the worldvolume theory. For this reason, there is nothing more
to learn for us from the 3--form equations of motion and we refer to
ref.~\cite{aaf} for further details.
 
\vspace{0.4cm}

To conclude, we have verified that all covariant background field
quantities and their transformations under $SL(2)\times SL(3)$ in
$D=8$, which we could ``reasonably'' read off from the membrane worldvolume
theory, are indeed correct, as they are in agreement with the
$SL(2)\times SL(3)$ invariance of the low energy effective theory.

\section{Discussion}

In this paper, we have attempted to derive U--duality symmetries as symmetries
of the membrane worldvolume theory. Our method was to rewrite the membrane
equations in a manifest covariant form by pairing equations of motion and
Bianchi identities of the original and the dual theory. In doing so, we
followed the route which leads to the discovery of T duality as a symmetry
of the string worldsheet as closely as possible.

For pure moduli backgrounds and dimensions $D>6$, we could derive the correct
U--duality group and moduli coset parameterization in such a way. This
generalizes the work of Duff and Lu~\cite{dl} and shows that their results can
actually be obtained by taking the full 11--dimensional target space
into account. Generally, manifest U duality in a dimensionally reduced
theory requires Poincar\'e dualization of certain background form fields.
For $D\leq 6$ this Poincar\'e dualization, which we do no know how to
explicitly carry out in the worldvolume theory, affects the moduli
sector of the membrane equations of motion. Therefore, we could not extend
our analysis for pure moduli backgrounds to lower dimensions, $D\leq 6$.

If all background fields are included, the need to Poincar\'e dualize also
influences the equations of motion for $D=7,8$. The cleanest, nontrivial
case in this respect is the one for $D=8$ with U--duality group
$SL(2)\times SL(3)$, since only the 3 form is affected.
For our general analysis, we therefore concentrated on this case. It turned
out that the membrane equations of motion could ``almost'' be written
in an $SL(2)\times SL(3)$--invariant form. Moreover, we were able to read off
the correct $SL(2)\times SL(3)$--covariant background field quantities for
all fields except the 3 form. These results have been verified by comparison
with $D=8$ supergravity obtained by dimensional reduction of $\DH =11$
supergravity. However, in each part of the membrane equations of motion
the $SL(2)$ part of the symmetry is obstructed by a term bilinear in the
external target space coordinates $X^\m$. Unfortunately, we have no
understanding why these terms appear in the internal and mixed equations
of motion. Since the symmetry breaking term in the external equations of
motion is associated with the 3 form, its origin is possibly related to
our ignorance of how to perform the Gaillard--Zumino construction for
the 3 form on the worldvolume. It is conceivable that the origin
of the other terms is similar. Perhaps this construction cannot be carried
out within the membrane theory, but only by
combining it with the 5--brane worldvolume theory (which, after all, contains
a 6 form dual to the membrane 3 form). This would imply that $SL(2)$ is not
a symmetry of the classical membrane worldvolume theory. Clearly, we are
not drawing such a conclusion from our results, since there are other possible
sources of symmetry breaking within our approach. In section 3, we have
pointed out that our solution~\refs{m_Vsol} of the dual theory is actually
not the most general one. We have, unsuccessfully, attempted to use the
remaining freedom in order to remove the obstruction. Still, we cannot
exclude the possibility that there exists a more general solution of
eq.~\refs{m_solv1} which leads to fully $SL(2)$ covariant equations of
motion. Another problem, pointed out in ref.~\cite{sez}, arises once 
operators bilinear in target space coordinates are split up into pieces. For
example, the operators $\cF^{(1)im} = \F^{-k}\sg\g^{ij}\pa_jX^m$ and
$\wtd{\cF}^{(2)i\mb}=\e^{ijk}\pa_jX^{m_1}\pa_kX^{m_2}$ transform as
upper and lower component of an $SL(2)$ doublet, though this seems to
contradict the fact that one is basically the square of the other. A related
problem arises, once the transformation of the worldvolume metric $\g_{ij}$
is taken into account. It can be computed from eq.~\refs{indmetric1} and the
various transformations for the vertex operators and the metric components.
This leads to a complicated nonlinear transformation of $\g_{ij}$ which seems
to be incompatible with the linear transformation of $\cF^{(1)im}$ and
$\wtd{\cF}^{(2)i\mb}$. Though these are important issues, which have to be
clarified, we feel that they should not be taken too serious. After all,
the charges associated with the conserved currents
$\left( \frac{\pa{\cal L}}{\pa (\pa_iX^m)}=\cF^{(1)im}\gh_{mn}+...,
\wtd{\cF}^{(2)i\mb}\right)$ do transform as a doublet under $SL(2)$~\cite{ss}.
And, perhaps even more significantly, despite the obstruction we encounter,
we are able to reproduce a large part of the U--duality symmetry structure
directly from the worldvolume. We hope that this is a step forward toward a
rigorous proof of U--dualities as symmetries of the membrane worldvolume.

\vspace{0.4cm}

{\bf Acknowledgments} We would like to thank D. Waldram for numerous
interesting discussions and M. Duff for a very helpful email conversation.
A.~L.~is supported by a fellowship from Deutsche Forschungsgemeinschaft (DFG).
A.~L.~and B.~A.~O.~are supported in part by DOE under contract
No.~DE-AC02-76-ER-03071.
\end{document}